\documentclass[ALICE,manyauthors]{cernphprep}
\usepackage[comma,square,numbers,sort&compress]{natbib}
\usepackage[utf8]{inputenc}
\usepackage{hyperref}
\usepackage{color}
\usepackage{lineno}
\usepackage{booktabs}
\usepackage{cernunits}
\usepackage[utf8]{inputenc}
\usepackage{multirow}
\usepackage{graphicx}
\usepackage[loose]{units}
\usepackage{upgreek}
\usepackage{amsmath,amssymb,amsbsy,mathrsfs} 
\usepackage[T1]{fontenc}
\usepackage{placeins}

\usepackage{color}
\usepackage[usenames,dvipsnames,svgnames]{xcolor}
    \definecolor{CERNblue}{RGB}{0,83,161}
    \definecolor{aliceRed}{RGB}{227,6,19}
\usepackage{tikz}
\usetikzlibrary{patterns, positioning, decorations.pathreplacing, decorations.pathmorphing, decorations.markings, calc, arrows, trees, through}
\usepackage{sansmath}
\usepackage{bbold}
\begin{document}%

\newcommand{\PbPb}{\textnormal{Pb--Pb}}
\newcommand{\AuAu}{\textnormal{Au--Au}}
\newcommand{\pA}{\textnormal{p--A}}
\newcommand{\pp}{\ensuremath{\mbox{p}\mbox{p}}}
\newcommand{\pt}{\ensuremath{p_\mathrm{T}}}
\newcommand{\pT}{\pt}
\newcommand{\etg}{\ensuremath{E_\mathrm{T}^{\gamma}}}
\newcommand{\ptg}{\ensuremath{p_\mathrm{T}^{\gamma}}}
\newcommand{\ETG}{\etg}
\newcommand{\PTG}{\ptg}
\newcommand{\GeVc}{\ensuremath{\mathrm{GeV}/c}}
\newcommand{\ptch}{\ensuremath{p_\mathrm{T,\,ch\;jet}}}
\newcommand{\deltaptch}{\ensuremath{\delta p_\mathrm{T,\,ch}}}
\newcommand {\snnbf}{\ensuremath{\mathbf{\sqrt{s_{_{\mathrm{NN}}}} }}}
\newcommand {\snn}{\ensuremath{\sqrt{s_{_{\mathrm{NN}}}} }}
\newcommand{\shshlo}{\ensuremath{\sigma_\mathrm{long}^{2}}}
\newcommand{\piz}{\ensuremath{\pi^{0}}}

\newcommand{\dirg}       {\ensuremath{{\rm \gamma, dir}~}}
\newcommand{\decg}       {\ensuremath{{\rm \gamma, dec}~}}
\newcommand{\incg}       {\ensuremath{{\rm \gamma, inc}~}}

\begin{titlepage}
\PHyear{2019}
\PHnumber{117}      
\PHdate{04 June}   

%

\title{Measurement of the inclusive isolated photon production cross section in $\pp$ collisions at  $\sqrt{s}$~=~7~TeV} 
\ShortTitle{Isolated photon production in ALICE}   

\Collaboration{ALICE Collaboration\thanks{See Appendix~\ref{app:collab} for the 
list of collaboration members}}
\ShortAuthor{ALICE Collaboration} 

\begin{abstract}
The production cross section of inclusive isolated photons has been measured by the ALICE experiment 
at the CERN LHC in pp collisions at a centre-of-momentum energy of $\sqrt{s}=$~7~TeV. The measurement is performed with the electromagnetic calorimeter EMCal and the central tracking
detectors, covering a range of $|\eta|<0.27$ in pseudorapidity 
and a transverse momentum range of $ 10 < \ptg < 60~\GeVc$. 
The result 
extends the \pt\ coverage of previously published results of the ATLAS and CMS experiments at the same collision energy to smaller \pt.

The measurement is compared to next-to-leading order perturbative QCD calculations 
and to the results from the ATLAS and CMS experiments.
All measurements and theory predictions are in agreement with each other.

\end{abstract}
\end{titlepage}

\setcounter{page}{2}
\section{Introduction}

In high-energy particle collisions, direct photons are those photons which are directly produced in elementary processes, and as such are not products from hadronic decays.
In proton-proton and nuclear collisions, direct photons are colourless probes of QCD processes. 
Photons originating from hard scatterings of partons from the incoming hadrons are called prompt photons. They provide a handle for 
testing perturbative QCD (pQCD) predictions, and they are probes of the initial state of protons 
or nuclei. At the lowest order (LO) in pQCD, prompt photons are produced via two processes: 
(i) quark-gluon Compton scattering, $q g \rightarrow q \gamma$,  (ii) quark-antiquark annihilation, $q \bar{q} \rightarrow g \gamma$, and, with a much smaller contribution, $q \bar{q} \rightarrow \gamma \gamma$. In addition, prompt photons are produced by higher-order processes, 
like fragmentation or bremsstrahlung \cite{Aurenche:1993}.
The collinear part of such processes has been shown to contribute effectively also at LO.

A discussion of early prompt photon measurements can be found in \cite{Ferbel:1984ef}, and measurements are also available from experiments at the SPS collider \cite{Appel:1986ix}, the Tevatron \cite{Abazov:2005wc,Aaltonen:2009ty} and RHIC \cite{PhysRevLett.98.012002}. Recently, measurements have been performed at the LHC by the ATLAS and CMS collaborations in pp collisions at various energies \cite{Khachatryan:2010fm,CMS2011,Chatrchyan:2012vq,Sirunyan2019, Aad:2010sp,Aad:2011tw,Aad:2013zba,Aad:2016xcr,Aad:2017}. %

These measurements allow one to study a wide range of transverse momentum (\pt) of prompt photon production from 15 to 1000 $\mathrm{GeV}/c$, the lowest limit being partially defined by the use of a high-energy photon trigger. A more fundamental limitation for direct photon measurements is imposed by the general experimental conditions. In particular, photon conversions in detector material imply a worsening of momentum resolution and signal reduction that is especially important at  low momentum.
 For converted photons, the original  energy may even be recovered for very high momentum, but a strong bias will be introduced at low transverse momentum. 
The low material budget in the ALICE experiment ($X/X_0=0.7-0.9$ in front of the photon detector) makes photon measurements at low \pt\ more reliable and allows one to move the \ptg~reach to a lower value.  

In some of the above-mentioned references, 
the term ``direct prompt photons'' is introduced to denote photons from the 
$2 \rightarrow 2$ processes and is contrasted in particular with fragmentation  or bremsstrahlung photons emitted directly from partons. 
We follow a nomenclature that was adopted in a CERN Yellow Report \cite{Arleo:PhotonLHCYellow} where direct photons referred to all photons that do not originate from hadronic decays and prompt photons to all photons that are directly emerging from a hard process or produced by bremsstrahlung, in any order of perturbative QCD. When needed, we speak explicitly of “photons from $2 \rightarrow 2$  processes” in this paper.

Photons from $2 \rightarrow 2$ processes provide clear constraints of the underlying parton kinematics, 
but making a clean separation between the different types of prompt photons is difficult. 

In a consistent theoretical description, the separate treatment of certain diagrams is somewhat arbitrary and only justified quantitatively to reach a desired accuracy in a given calculation. A physical definition of a subset of photons has to be related to measurable criteria.
This has led to the prescription of so-called 
``isolated photons''. An isolation criterion is applied on photon candidates, where one requires 
the sum of the transverse energies (or transverse momenta)
of produced particles in a cone around the photon direction to be smaller than 
a given threshold value -- this can be done both in the experiment and in theoretical calculations. 
Fragmentation and bremsstrahlung photons are expected to be accompanied by
fragments of the parton that has been close in phase space, while photons from $2 \rightarrow 2$ processes should 
be free of such associated fragments.   
Thus, an isolation cut 
should significantly suppress fragmentation and bremsstrahlung, while it should affect the 
$2 \rightarrow 2$ processes only marginally \cite{Ichou:2010wc}.
A strong additional motivation of an isolation cut is to reduce the background of decay photons in the measured signal. This can be achieved, because hadrons at reasonably high \pt\ would in general be produced in jet fragmentation and would thus be accompanied in their vicinity by other jet fragments.

Measurements of prompt and in particular isolated photons provide constraints on the proton \cite{Ichou:2010wc} and nuclear \cite{Arleo:2011gc} Parton Distribution Functions (PDFs). At the LHC, because of the high centre-of-momentum energy ($\sqrt s$), 
such PDF studies are potentially sensitive to very small values of the longitudinal momentum fraction $x$ of the initial-state parton. For a $2 \rightarrow 2$ process with the two 
particles (3 and 4) in the final state being emitted at similar
rapidities $y_3\approx y_4 \approx y$, which is the dominant contribution to the inclusive single particle cross section,
the $x$ values in the initial state can to a good approximation be calculated as:
\begin{equation}
x_{1,2} \approx \frac{2 \pt}{\sqrt{s}} \exp(\pm y) \equiv x_{\mathrm{T}} \exp(\pm y),
\label{eq-bjorkenx}
\end{equation}
where \pt\ is the transverse momentum and $y$ the rapidity of the final state particles.
The variable $x_{\mathrm{T}}$ defined here is often used to compare transverse momentum distributions for different beam energies. For photons measured at mid-rapidity ($y \approx 0$), it is closely related to the Bjorken $x$-values: $x_{\mathrm{T}} \approx x_1 \approx x_2 \equiv x$.
At the LHC, the most important contribution to photon production, the quark-gluon Compton diagram, where the above relation can be applied, has the additional advantage to be directly sensitive to the gluon density, which has the largest uncertainty among the PDFs. 
These $2 \rightarrow 2$ processes, which should be enriched in the measurement via the isolation cut, should therefore probe the low-$x$ gluon content of one of the incoming hadrons. Any higher order effects will weaken the kinematic constraints, and in particular, fragmentation photons will be sensitive to much larger values of $x$. Also, when measuring only one of the final-state particles, the uncertainty on the rapidity of the other particles will lead to a broadening of the distribution of $x$ values probed. 

\begin{figure}[ht]
\begin{center}
\includegraphics[width=1.\textwidth]{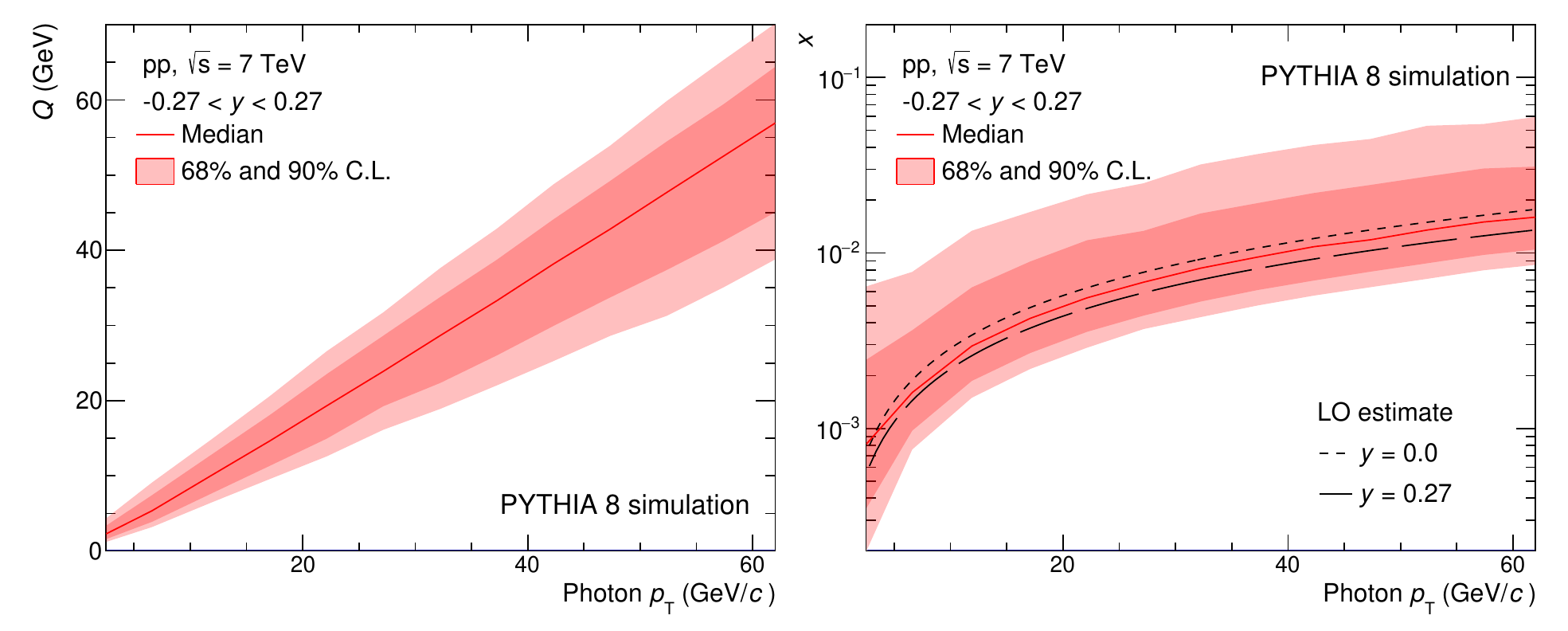} 
\end{center}
\caption{\label{fig:xPDF} 
(colour online) The scale $Q$ (left), and the fraction $x$ (right) of longitudinal momentum of the initial state partons of the hard process for photon production at midrapidity versus photon \pt\ for pp collisions at $\sqrt s = 7$~TeV, from a PYTHIA 8 \cite{Sjostrand:2014zea} simulation. The $x$ values of both partons are used here. The solid red line shows the median of the distribution in the parameters, while the red bands indicate the 68\% and 90\% confidence level intervals. The right plot shows for comparison the LO estimate for $x$ according to Eq.~\eqref{eq-bjorkenx} for the rapidity range studied here. 
}
\end{figure}

To get a better understanding of the ranges of kinematic parameters of the partonic processes that are explored in prompt photon measurements, we have performed a study of photon production with the PYTHIA 8 generator (version 8.235 \cite{Sjostrand:2014zea} with Monash 2013 Tune \cite{Skands:2014pea}), where we extract the values of factorisation scale $Q$ and parton momentum fractions $x_1$ and $x_2$ directly as specified in the PYTHIA event record.
PYTHIA does not contain all effects of higher orders in QCD systematically, but has some important enhancements beyond pure LO via initial- and final-state radiation. 
In this calculation, we have not implemented an isolation cut, but we assume that for the purpose of this study, it is equivalent to simulating only the LO processes for partonic photon production. 
Fig.~\ref{fig:xPDF} shows the results of this calculation. 

The value of the scale $Q$ (left panel) 
is approximately proportional to {\pt}. 
The momentum fractions $x$ of the two partons are shown in the right panel. Here, we include both partons, because for a midrapidity measurement $x_1$ and $x_2$ should be approximately equivalent. 
For comparison, the LO estimates from Eq.~\eqref{eq-bjorkenx} are also shown in the figure. The overall dependence of the values of $x$ on \pt\ is very similar, and they 
show the expected increase with {\pt}.  

These results suggest that, for most of the \pt\ range for a photon measurement at midrapidity, the LO estimate is a reasonably good description of the overall behaviour of the kinematics. However, there is a significant width to the distributions. For example, while a photon of  $\pt = 10 \,\  \mathrm{GeV}/c$ should be dominantly sensitive to values of $x \approx 2 \cdot 10^{-3}$, 
there should be significant contributions up to $x \approx 10^{-2}$. 
Similar behaviour is expected for other probes, e.g.\ hadrons, but the latter will have a larger spread in their kinematic sensitivity and will in general probe larger values of $x$ due to fragmentation, and they will thus be less selective in terms of the kinematics. 
This behaviour motivates the measurement of isolated photons at the lowest possible values of \pt. 

Direct photon measurements can provide additional information in high-energy nuclear collisions.
Prompt photon measurements in nucleus-nucleus collisions can yield a 
reference for the medium-induced modification of strongly-interacting probes. 
In particular, jets and high transverse momentum hadrons are suppressed in such 
collisions \cite{ALICERAApi0,ALICERAACh,ALICEJetSupp}, while photons, similar to W and $\text{Z}^0$ bosons, should be unaffected by the strongly interacting medium,
consistent with the first  measurements of prompt photons in AA collisions \cite{Chatrchyan:2012vq,Aad2016PbPb}. 
Measurements in pp collisions provide a baseline for the latter.

At lower \pt, other sources of direct photons than prompt photons exist, in particular in high-energy nuclear collisions, 
where e.g.\ thermal photons are produced from the thermalised system.
These are important probes of the quark-gluon plasma. Their contribution has been measured by experiments at the SPS \cite{Aggarwal:2000th}, RHIC \cite{Afanasiev:2012dg}, and the LHC \cite{Adam:2015lda}.

In this paper, results of the ALICE experiment on isolated photon measurements in pp collisions at $\sqrt{s} = 7 \,\mathrm{TeV}$ are presented.  
The paper is organised as follows: Section~\ref{sec:detector} presents the detector setup and data sample analysed, the analysis procedure and uncertainties are described in Sect.~\ref{sec:analysis} and~\ref{sec:sys_unc} and finally, results and conclusions are presented in Sect.~\ref{sec:results} and~\ref{sec:conclusion}, respectively.

\section{Detector description and data selection}
\label{sec:detector}

A comprehensive description of the ALICE experiment and its performance is provided 
in Refs. \cite{Aamodt:2008zz,Abelev:2014ffa}.
Photon reconstruction was performed using the Electromagnetic 
Calorimeter (EMCal) \cite{Cortese:2008zza} while charged particles are reconstructed with the combination of the Inner Tracking System (ITS)~\cite{Aamodt:2010aa} and the Time Projection Chamber (TPC)~\cite{Alme:2010ke}, which are part of the ALICE central tracking detectors.

The ITS consists of six layers of silicon detectors and surrounds the interaction point, covering full azimuth. The two innermost layers consist of the Silicon Pixel Detector (SPD) and are positioned at radial distances of \unit[3.9]{cm} and \unit[7.6]{cm}. They are surrounded by the two layers of the Silicon Drift Detector (SDD) at \unit[15.0]{cm} and \unit[23.9]{cm}, and by those of the Silicon Strip Detector (SSD) at \unit[38.0]{cm} and \unit[43.0]{cm}. 
While the two SPD layers cover $|\eta| <$ 2 and $|\eta| <$ 1.4, respectively, the SDD and the SSD subtend $|\eta| <$ 0.9 
and $|\eta| <$ 1.0, respectively.
The TPC is a large ($\approx$ \unit[85]{m$^3$}) cylindrical drift detector filled  with a
Ne/CO$_2$ gas mixture. It covers $|\eta| <$ 0.9 over the full
azimuth angle, with a maximum of 159 reconstructed space points along the track path. 
The TPC and ITS tracking points are matched when possible, forming tracks with an associated momentum.

The EMCal is a lead-scintillator sampling electromagnetic calorimeter used to measure photons, electrons and the neutral part of jets via the electromagnetic showers that the different particles produce in cells of the calorimeter.
The scintillating light is collected by optical fibres coupled to Avalanche Photo Diodes 
(APD) that amplify the signal.
The energy resolution is $\sigma_E/E = A\oplus B/\sqrt{E}\oplus C/E$
with  $A = (1.7 \pm 0.3) \%$, $B = (11.3 \pm 0.5) \%$, $C = (4.8 \pm  0.8) \%$ and
energy $E$ in units of GeV~\cite{Abeysekara:2010ze}.
The EMCal was installed at a radial distance of \unit[4.28]{m} from the ALICE interaction point.
During the period in which the analysed dataset was collected, the EMCal consisted of 10 supermodules 
with a total aperture of $|\eta|<0.7$ in pseudorapidity and $80^\circ < \varphi < 180^\circ$ in azimuthal angle. 
The supermodules are subdivided into $24 \times 48$ cells, 
each cell with transverse size  of $6 \times 6~\mbox{cm}^2$ which corresponds to $\Delta \eta \times \Delta \varphi $ = 0.0143$\times$0.0143 rad, approximately twice the Moli\`ere radius. Thus, most of the energy of a single photon is deposited in a single cell plus some adjacent ones. 
The minimum bias interaction trigger was based on the response of the V0 detector, consisting of two arrays of 32 plastic scintillators, located at $2.8 < \eta < 5.1$ (V0A) 
and  $-3.7 < \eta < -1.7$ (V0C) \cite{Abbas:2013taa}.

The data used for the present analysis were collected during the 2011 LHC data taking period with 
pp collisions at the centre-of-momentum energy of $\sqrt s = 7$~TeV.  The analysed data were selected 
by the EMCal Level-0 (L0) trigger requiring energy deposition larger than 5.5~GeV in a tile of 2$\times$2 adjacent cells, 
in addition to the Minimum Bias trigger condition (MB, a hit in either V0A, V0C or SPD). 
The L0 decision, issued  at latest 1.2 µs after the collision, is based on the analog charge sum of the cell tiles
evaluated with a sliding window algorithm within each physical 
Trigger Region Unit (TRU) spanning $4\times 24$ cells.

The integrated luminosity 
taken with the EMCal trigger ($\mathscr{L}$) 
has been determined using the expression

\begin{equation}
\mathscr{L} = \frac{N_{\rm evt}~R} { \sigma_{\rm MB}} 
\end{equation}
where $N_{\rm evt}= 8.85 \cdot 10^6$
is the number of events selected with the EMCal trigger and $\sigma_{\rm MB}=53.7 \pm 2.0$~mb \cite{ALICE:INT7CrossSection} the measured minimum bias trigger cross section for the year 2011 sample.

Furthermore, $R$ is the trigger rejection factor, which quantifies the fraction of interaction triggers which are rejected by the additional EMCal L0 trigger condition. It has been corrected for in-bunch pile-up (average number of collisions per bunch crossing) and amounts to $R = 2941\pm 174$.
The resulting sampled luminosity of the current measurement is
$\mathscr{L}=473\pm 28$~(stat)~$\pm 17$~(syst)~$\mbox{nb}^{-1}$.

\section{Isolated photon reconstruction and corrections}
\label{sec:analysis}
Direct photon identification used in this analysis is based on three steps: 
 (a) particle reconstruction in the calorimeter;
 (b) photon identification via track-cluster matching cuts and the study of the shower shape produced by 
 the particle; 
 and (c) selection of isolated photons.

The detector response is modelled by Monte Carlo (MC) simulations reproducing the same detector conditions as for the data taking period.
The corrections discussed  in the next subsections  are obtained using PYTHIA 6 (version 6.421~\cite{Sjostrand2006}, with Perugia 2011 tune \cite{PerugiaTune} and CTEQ5L for PDF \cite{CTEQ5} ) as particle generator simulations, generating processes in bins of transverse momentum of the hard scattering 
with two jets (jet-jet) or a direct photon and a jet 
($\gamma$-jet, mainly Compton and annihilation processes) as final state, 
and GEANT3~\cite{Geant3} for particle 
transport in the detector material. 
In the case of $\gamma$-jet event generation, the event 
is accepted when the direct photon enters the EMCal acceptance. 
In the case of jet-jet event generation, the event is accepted when at least one jet 
produces a high-\pt\ photon originating from a hadron decay in the EMCal acceptance. 
To enhance the number of such photons, which are the main 
background in this analysis, two sub-samples  have been used in the jet-jet simulation, each with different event selection, where it is ensured that a decay photon with  $\pt > 3.5$ or {7~\GeVc} is present in the EMCal acceptance.

\subsection{Cluster reconstruction and selection \label{sec:clust}}
Particles deposit their energy in several calorimeter cells, forming a cluster. 
Clusters are obtained by grouping all cells with common sides whose energy
is above 100\,MeV, starting from a seed cell with at least 300\,MeV. Furthermore, a cluster 
must contain at least two cells to ensure a minimum cluster size and to remove 
single-cell electronic noise fluctuations. 
In order to limit energy leakage at the supermodule borders, a distance of at least one cell 
of the highest-energy cell in the cluster to the supermodule border is required. These requirements lead to an acceptance of  $|\eta|<0.67$ in pseudorapidity and $82^\circ < \varphi < 178^\circ$ in azimuth.
During the 2011 data taking period, the LHC delivered events in bunches separated by 50\,ns. Therefore,
to ensure the selection of clusters from the main bunch crossing, the timing of 
the highest-energy cell in the clusters relative to the main bunch crossing has to satisfy $\Delta t<30$~ns. 

Finally, an energy non-linearity correction derived from electron test beam data  \cite{ALLEN20106}, of about 7\% at 0.5\,GeV and negligible above 3\,GeV, is applied to the reconstructed cluster energy. 

Nuclear interactions occurring in the APD, in particular those involving neutrons, 
induce an abnormal signal~\cite{Bialas:2013wra}. 
Such a signal is most frequently observed as a single 
high-energy cell with a few surrounding low-energy cells, and can be removed by
comparing the amplitudes in adjacent cells to the cell with maximum amplitude $E_{\rm max}$. To reject these signals, one requires that the ratio $F_{+} \equiv 1-E_{+}/E_{\rm max}$, where $E_{+}$ is the sum of the amplitude of the four surrounding cells that share a common edge with the maximum cell, satisfies $F_{+} < 97$\%.

Contamination of the cluster sample by charged particles is suppressed by a charged particle veto (CPV). 
It is provided by TPC tracks constrained to the vertex, selected so that the 
distance of closest approach 
to the primary vertex is less than 2.4\,cm in the plane transverse to the beam, 
and less than 3.0\,cm in the beam direction. 
The separation of the position of the track extrapolated to the EMCal surface from the cluster position must fulfil the conditions 
\begin{equation}
 \Delta \eta^\text{residual} > 0.010 +(\pt^\text{track} +4.07)^{-2.5}\ \mathrm{and}\ \Delta \varphi^\text{residual} > 0.015 +(\pt^\text{track} +3.65)^{-2} ~\mathrm{rad}
\end{equation}

where $\Delta \varphi^\text{residual}=|\varphi^\text{track}-\varphi^\text{cluster}|$, 
$\Delta \eta^\text{residual}=|\eta^\text{track}-\eta^\text{cluster}|$ 
and the track transverse momentum ($\pt^\text{track}$) is in \GeVc\ units as detailed in \cite{Achariya:2017}.
The track-to-cluster matching efficiency amounts to about 92\% for primary charged hadrons and electrons at cluster energies of $E \simeq 1$\,GeV, up to 96\% for clusters of 10\,GeV. 

From now on, clusters that pass the previous selection cuts are called ``neutral clusters''.

\subsection{Shower shape and photon identification}
\label{sec:photonident}
The neutral cluster can have a wider shape, if one or several additional 
particles deposit their energy nearby in the detector. The most frequent case is a two-particle cluster that has an elongated shape.
If the distance between particles 
is larger than two cells, one can observe cells with local maxima in the energy distribution 
of the cluster, where a local maximum is defined as a cell with a signal higher than 
the neighbouring cells. 

For an increasing number of local maxima ($N_{\rm LM}$),
the cluster will in general get wider. 
Direct photons generate clusters with
$N_{\rm LM}=1$, except if they suffer conversion in the material in front of the EMCal. 
The two decay photons from high-\pt\ $\piz$ and $\eta$ mesons with energy 
above 6 and 24\,GeV, respectively, likely merge into a single cluster as observed in simulations.
Merged clusters from $\piz$ mesons below 15 GeV and $\eta$ mesons below 60\,GeV most often have 
$N_{\rm LM} = 2$. With increasing energy the two-photon opening angle decreases, leading to 
merged clusters with mainly $N_{\rm LM} = 1$ above 25\,GeV for $\piz$ mesons and above 100\,GeV for $\eta$ mesons. 

We reject clusters with $N_{\rm LM}>2$ in
this analysis, as these clusters are the major contribution to the background and
contributions from more than 2 particles are not perfectly reproduced in Monte-Carlo
simulations. Contribution of clusters with $N_{\rm LM}=2$ is especially large in case of wide showers,
and are crucial for the estimate of the contamination of the direct photon sample, as
explained in Sect.~\ref{sec:purity}.

Merged and single photon clusters can be discriminated based on the shower shape using the width parameter \shshlo, the square of the larger eigenvalue of the energy distribution in the $\eta-\varphi$ plane \cite{Abelev:2014ffa,Adam:2016,Achariya:2017}, that can be calculated as:

\begin{equation}
\label{eq:ss1}
\sigma_{\rm long}^{2} = (\sigma_{\varphi\varphi}^{2}+\sigma_{\eta\eta}^{2})/2+\sqrt{(\sigma_{\varphi\varphi}^{2}-\sigma_{\eta\eta}^{2})^{2} / 4+\sigma_{\eta\varphi}^{4}},\\
\end{equation}

where ${\sigma^2_{xz}} = \big \langle xz \big \rangle - \big \langle x \big \rangle \big \langle z \big \rangle$ and $\big \langle x \big \rangle = (1/w_{\rm tot}) \sum w_i x_i$ 
are weighted over all cells associated with the cluster in the $\varphi$ or $\eta$ direction. 

\begin{figure}[ht]
\begin{center}
\includegraphics[width=0.49\textwidth]{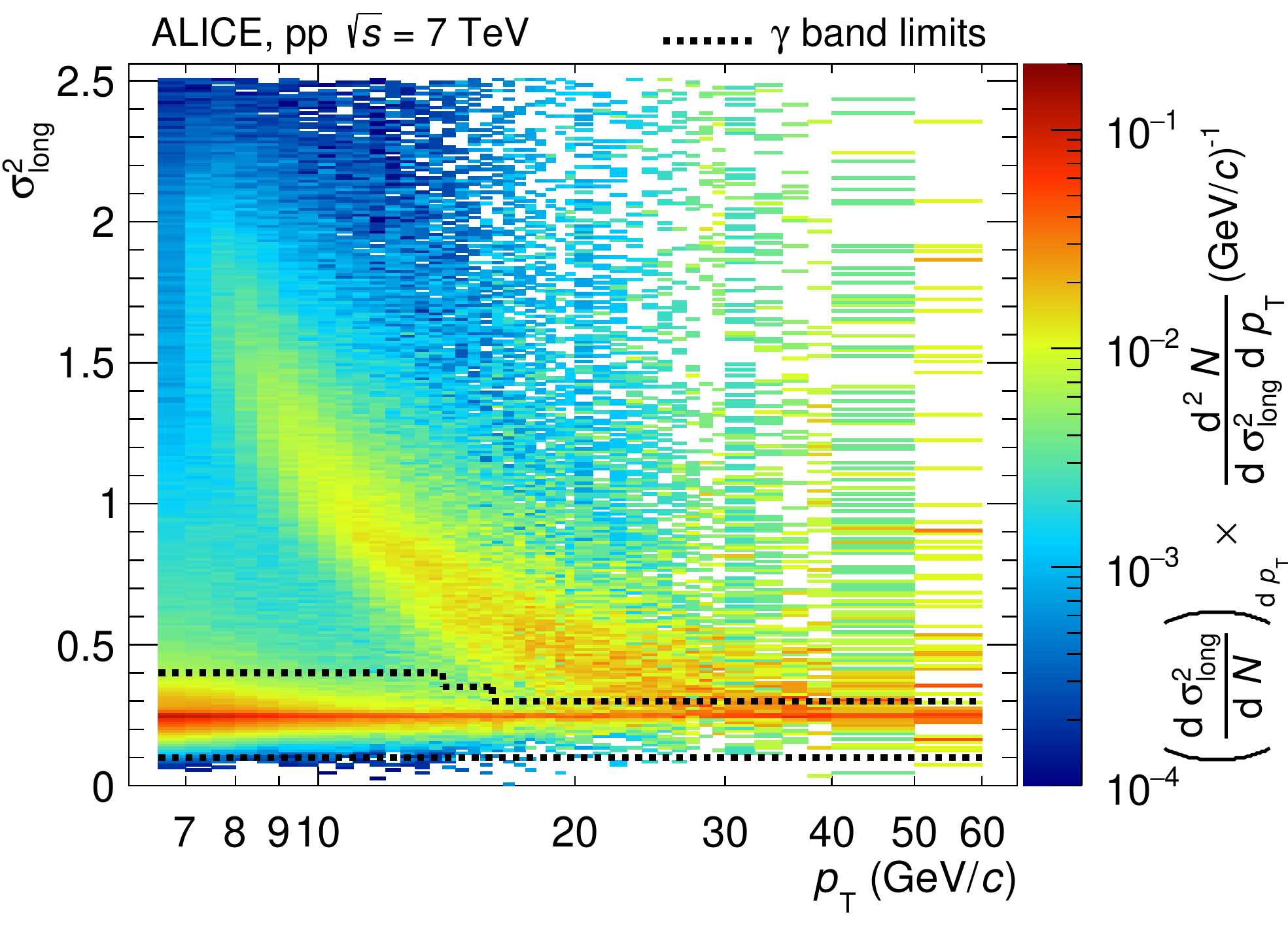} ~~~
\includegraphics[width=0.48\textwidth]{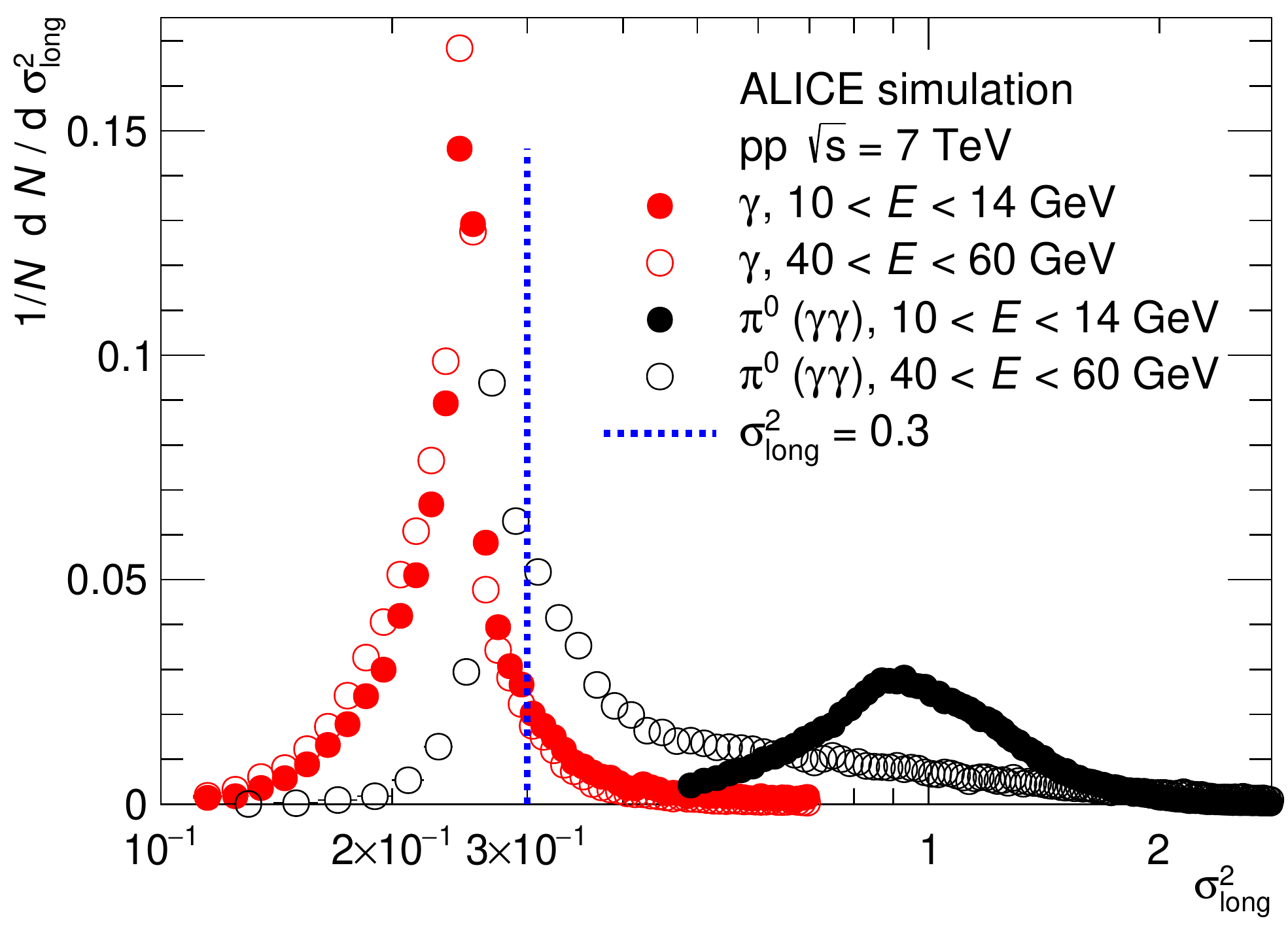}
\end{center}
\caption{\label{fig:SigmaLongExample} (color online) Left: Shower shape parameter \shshlo\ versus neutral cluster \pt.  
The band limited by the black dashed lines indicates the area populated by single photons and defines the photon selection cuts (narrow clusters) used in the analysis. 
Right: Distributions of the shower shape parameter
for different types of clusters produced by single photons (red bullets) or overlapped photons from \piz\ decays (black bullets) for different cluster energies from PYTHIA 6 $\gamma$-jet and jet-jet simulations with GEANT3 tuned for cross-talk emulation. 
A line at  $\shshlo=0.3$ represents the cut used to select narrow clusters at $\ptg > 18$\,\GeVc. 
All distributions are normalised to an integral of 1. In the left plot each \pt\ bin is separately normalised to 1.}
\end{figure}

The weights $w_i$ depend logarithmically on the ratio of the energy 
$E_i$ of the $i$-th cell 
to the cluster energy, as $w_i = \mathrm{max}(0,4.5+\ln(E_{i}/E_{\rm cluster}))$ and 
$w_{\rm tot} = \sum w_i$~\cite{Awes:1992}.
The neutral clusters \shshlo\ distributions as a function 
of \pt\ in data and simulation are shown in Fig.~\ref{fig:SigmaLongExample}. 
Most of the pure single photons are reconstructed as clusters with $\shshlo \approx 0.25$, 
other cases contribute to higher values as seen in simulation in Fig.~\ref{fig:SigmaLongExample}, right.
Above the higher limit in \shshlo\ (defined by the solid line), a clear \pt-dependent band 
populated by merged \piz-decay photons can be observed as shown in Fig.~\ref{fig:SigmaLongExample}, left.  
The value of \shshlo\ for merged decay photons decreases with increasing energy, which leads to an overlap with single photon showers. 
In this analysis, ``photon candidates'' refer to clusters with a narrow shape $0.1 < \shshlo < \sigma^2_{\rm max}$ with $\sigma^2_{\rm max} = 0.4$ in 
$10 < \ptg < 14$\,\GeVc, $\sigma^2_{\rm max} = 0.35$ in $14 < \ptg < 16$\,\GeVc\ 
and $\sigma^2_{\rm max} = 0.3$ for $\ptg > 16$\,\GeVc. 

\begin{figure}[ht]
\begin{center}

\includegraphics[width=0.49\textwidth]{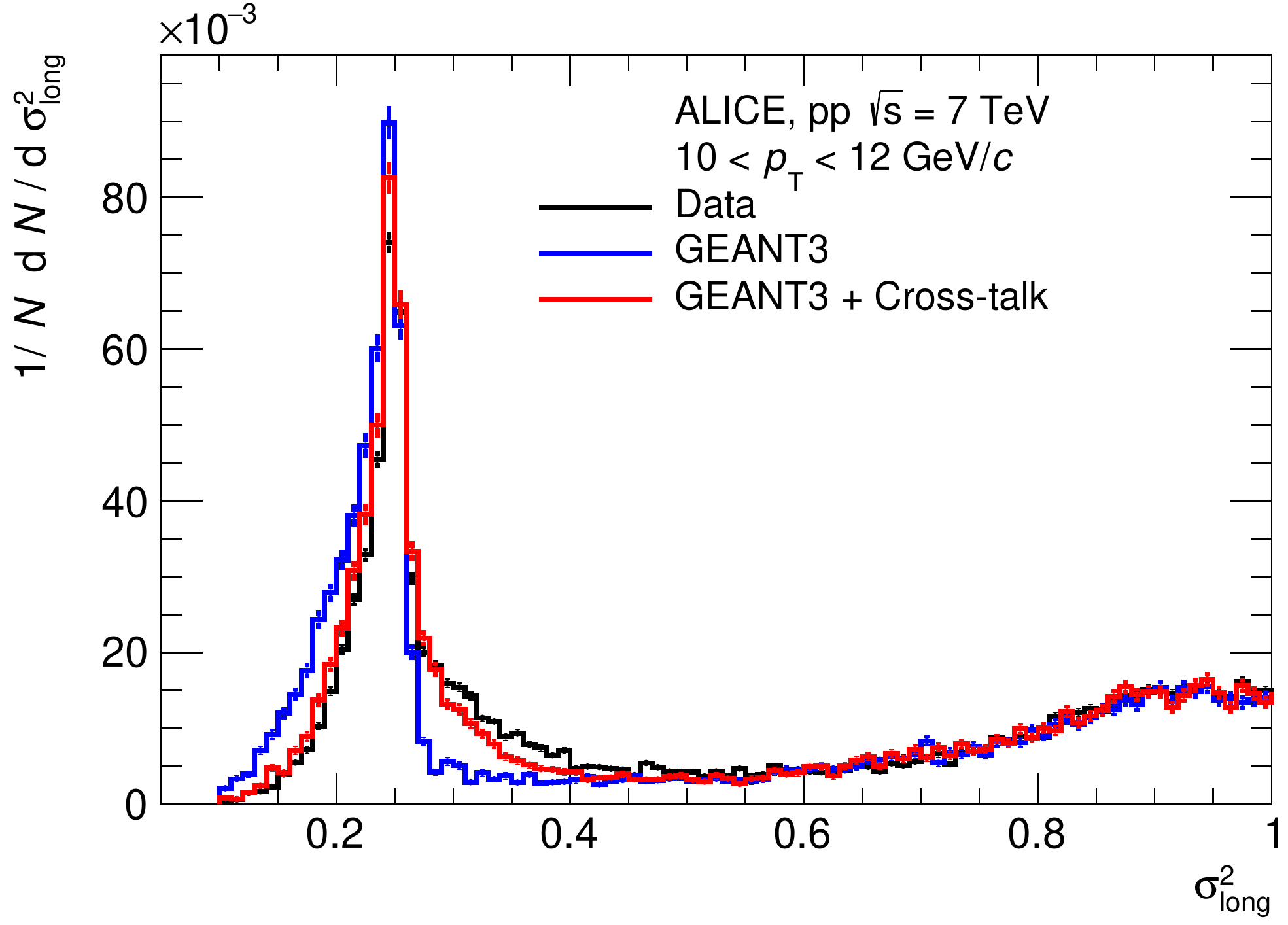}
\includegraphics[width=0.49\textwidth]{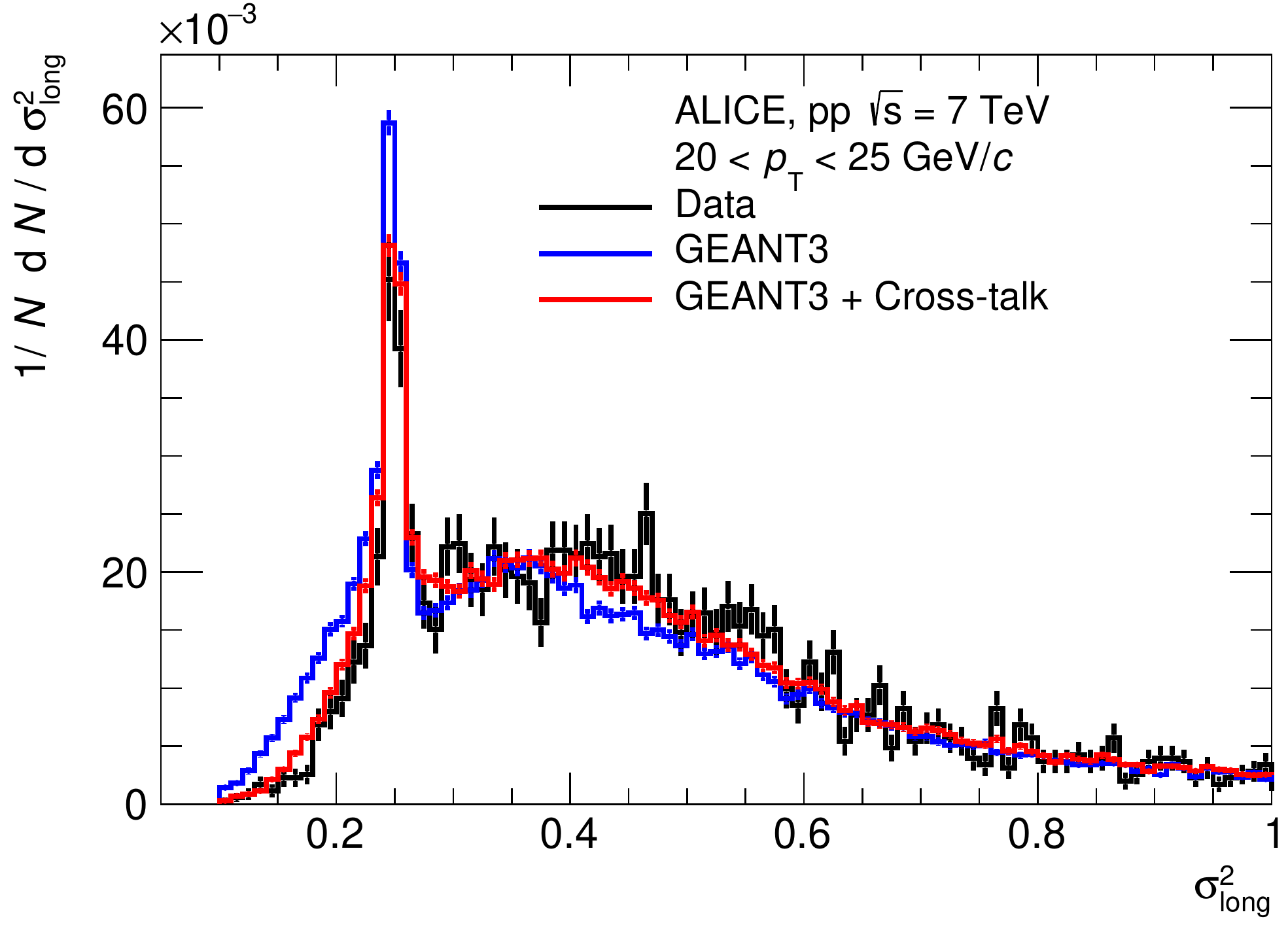}

\end{center}
\caption{\label{fig:SigmaLongDataMCxTalk} (colour online) Distributions of the shower shape parameter $\shshlo$ of neutral clusters in data and simulations as used in this analysis. The different panels show different neutral cluster \pt\ intervals. All distributions are normalised to the integral. Data are shown as black histograms and simulations, PYTHIA 6 jet-jet $+$ $\gamma$-jet events, with GEANT3 default settings in blue. The red histograms are also simulations based on GEANT3, but tuned to reproduce the cross-talk observed in the EMCal electronics.}
\end{figure}

A comparison of the shower shape parameter
\shshlo\ distribution in data and Monte-Carlo simulations is shown in Fig.~\ref{fig:SigmaLongDataMCxTalk}. The single photon peak
in data compared to simulations has a stronger tail towards larger values of \shshlo, i.e.
specifically in the region $0.3 \le \shshlo \le 0.4$.

The main reason for this was identified as a cross-talk between 
cells belonging to the same EMCal readout card, called T-Card, which serves $2\times8$ cells 
(in $\eta\times\varphi$). The cross-talk results in an increase of the amplitude in cells close 
to the highest-energy cell of the cluster and in the same T-Card, with a few percent 
of $E_{\rm max}$, which in turn leads to a modification of the cluster shape. This 
effect was modelled in the simulation and a good agreement between data and simulation 
was achieved, as seen in Fig.~\ref{fig:SigmaLongDataMCxTalk} for two 
neutral cluster \pt\ 
intervals.

\subsection{Isolated photon selection}
\label{sec:isolation}

Direct photons emitted in  $2 \rightarrow 2$ processes are mostly  isolated, i.e.\ have 
 no hadronic activity in their vicinity except for the underlying event of the collision, in contrast 
to other photon sources like photons from parton fragmentation or decays of hadrons which have a high probability to
be accompanied by other fragments
~\cite{Ichou:2010wc}.

An isolation criterion is applied to direct photon candidates to increase the purity of $2 \rightarrow 2$ processes. 
As a consequence for a comparison with theory, one has to make sure to implement an equivalent cut in the theoretical calculations.

The isolation criterion is based on the so-called ``isolation momentum'' $p_{\rm T}^{\rm iso}$, i.e. the 
transverse momentum of all particles measured inside a cone  
around the photon candidate, located at $\eta^{\gamma}$ and $\varphi^{\gamma}$. The cone radius used is defined as: 
\begin{equation}
\label{eq:rsize}
R = \sqrt{ (\eta - \eta^{\gamma})^2 + (\varphi-\varphi^{\gamma})^2}
\end{equation}

We chose $R=0.4$ as cone radius as it contains the dominant fraction of the jet energy \cite{PhysRevD.71.112002} and is sufficiently large to contain both decay products of neutral meson decays. In addition it is fully contained within the acceptance of the electromagnetic calorimeter.

The isolation momentum is the sum of the transverse momenta of all neutral clusters ($p_{\rm T}^{\rm cluster}$) in the calorimeter, excluding the candidate photon, and of the transverse momenta of all charged tracks that fall into the cone:
\begin{equation}
\label{eq:etiso}
p_{\rm T}^{\rm iso}=\sum p_{\rm T}^{\rm cluster}+\sum p_{\rm T}^{\rm track}.
\end{equation}
The candidate photon is declared isolated if $p_{\rm T}^{\rm iso}<$~2\,\GeVc. 
This value was chosen after studying the efficiency, background rejection and purity performances, and optimizing these quantities. 
In order to have full coverage of the cone in the calorimeter, the photon candidate is restricted to a fiducial acceptance corresponding to $|\eta^{\gamma}|<0.27$ in pseudorapidity and 
$103^{\circ}<\varphi^{\gamma} < 157^{\circ}$ in azimuth.

Charged particles used in the calculation of the isolation momentum are reconstructed in a hybrid approach using ITS and TPC, which reduces local inefficiencies potentially caused by non-functioning elements of the ITS.
Two distinct track classes are accepted in this method~\cite{Abelev:2014ffa}: (i) tracks containing at least three hits in the ITS, including at least one hit in the SPD, with momentum determined without the primary vertex constraint, and (ii) tracks containing less than three hits in the ITS or no hit in the SPD, with the primary vertex included in the momentum determination. Class (ii) is used only when layers of the SPD are inactive in the acceptance.
Class (i) contributes 90\% and class (ii) 10\% of all accepted tracks, independently of \pt. The same constraints to the vertex as for TPC tracks discussed before are required.
Accepted tracks satisfy $|\eta^{\rm track}|< 0.9$ and $\pt^{\rm track}>0.2$~\GeVc.

\subsection{Purity of the isolated photon sample\label{sec:purity}}

The isolated photon candidate sample still has a non-negligible contribution from background
clusters, mainly from neutral meson decay photons.  
To estimate the background contamination, different classes of measured clusters 
were used:
(1) classes based on the shower shape \shshlo, i.e.\ \textit{narrow}, photon-like and \textit{wide} (most often elongated, i.e.\ non-circular), and 
(2) classes defined by the isolation momentum $p_{\rm T}^{\rm iso}$, i.e.\ \textit{isolated} (iso) and \textit{non-isolated} ($\overline{\rm iso}$).  
The different classes are denoted 
by sub- and superscripts, e.g. isolated, narrow clusters are given as $X_{\rm n}^{\rm iso}$ 
and non-isolated, wide cluster are given as $X_{\rm w}^{\overline{\rm iso}}$. The  {\shshlo} parameter
values for narrow and wide clusters correspond to the signal and background  clusters indicated in Sect.~\ref{sec:photonident}. The wide clusters use $ 0.55 < \shshlo <1.75$ for $\ptg <$ 12 \GeVc, $ 0.5 < \shshlo <1.7$ for $12 < \ptg < 14$ \GeVc,  $ 0.45 <  \shshlo <1.65$ for $14 < \ptg < 16$ \GeVc\ and  $ 0.4 < \shshlo <1.6$ for $\ptg >$ 16 \GeVc. 

The isolation criterion corresponds to $p_{\rm T}^{\rm iso}<$~2\,\GeVc\ whereas the anti-isolation corresponds to $p_{\rm T}^{\rm iso}>$~3\,\GeVc. 
The yield of isolated photon candidates in this nomenclature is $N_{\rm n}^{\rm iso}$. 
It consists of signal ($S$) and background ($B$) contributions: 
$N_{\rm n}^{\rm iso} = S_{\rm n}^{\rm iso} + B_{\rm n}^{\rm iso}$.

\begin{figure}
\begin{center}
\includegraphics[width=0.5\textwidth]{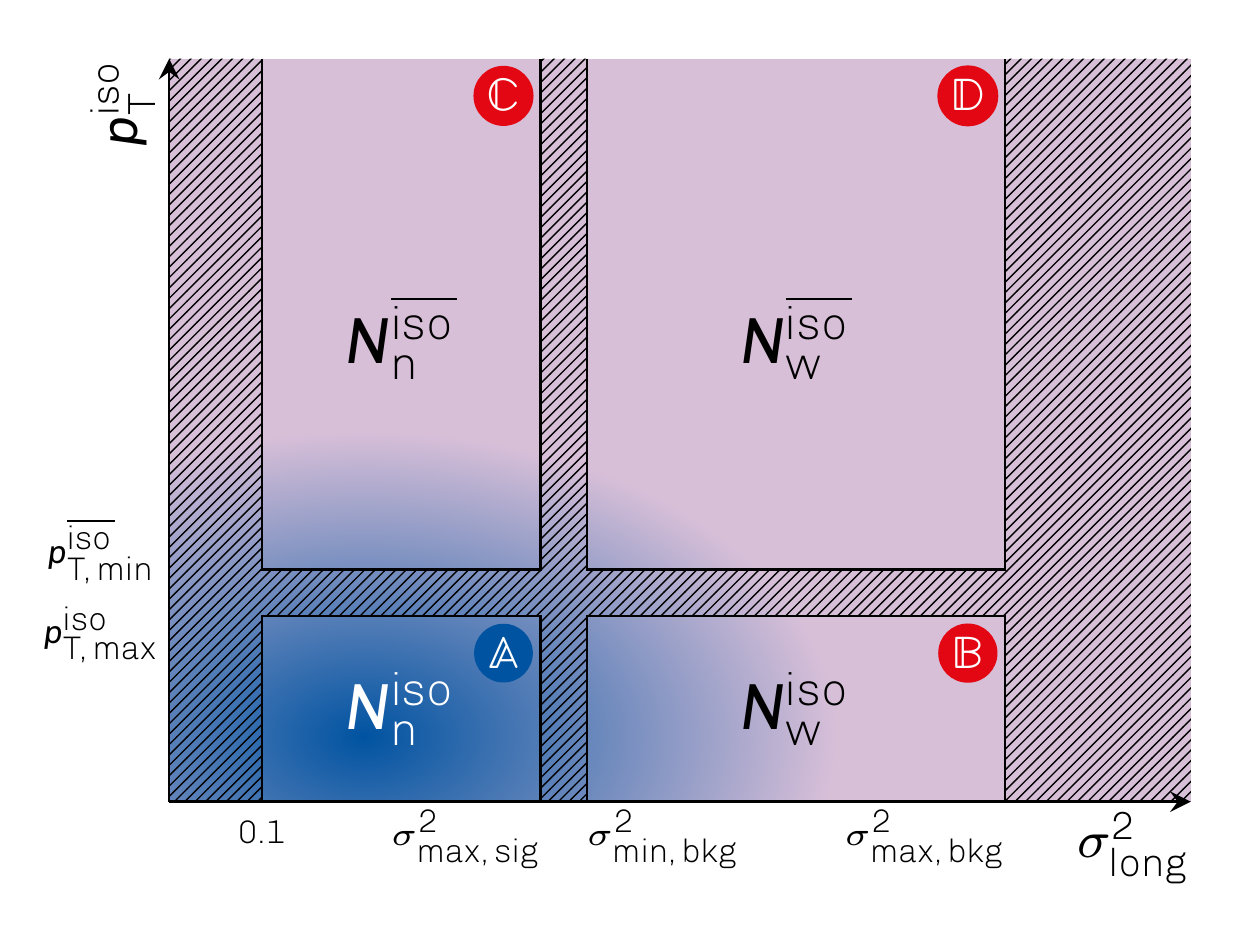} 
\end{center}
\caption{\label{fig:abcdIsoPhoton} (colour online) Illustration of the parametric-space of the photon isolation momentum and the shower width parameter (\shshlo), used to estimate the background yield in the signal region ($\mathbb{A}$) from the observed yields in the three control regions ($\mathbb{B}$, $\mathbb{C}$, $\mathbb{D}$). The red regions indicate areas dominated by background and the blue regions those that contain the photon signal. The colour gradient between these regions illustrates the presence of a signal contribution in the three background zones.}
\end{figure}

This class is labelled with the letter $\mathbb{A}$ in Fig.~\ref{fig:abcdIsoPhoton}, which illustrates the parameter 
space used in this procedure. The three other classes that can be defined 
(labelled as $\mathbb{B}$, $\mathbb{C}$, and $\mathbb{D}$ in the figure) should dominantly contain
background clusters. The notation $\mathbb{A}$, $\mathbb{B}$, $\mathbb{C}$ and $\mathbb{D}$ is analogous to the one used by the ATLAS experiment for their contamination estimate \cite{Aad:2011tw}. The contamination of the candidate sample is then $C = B_{\rm n}^{\rm iso}/N_{\rm n}^{\rm iso}$, or respectively, the purity is then $P \equiv 1 - C$.
Assuming that the ratios of isolated over non isolated  background in the narrow cluster areas 
is the same as in the wide cluster areas 
so that

\begin{equation}
\label{eq:ABCDproportionality}
\frac{B_{\rm n}^{\rm iso}/B_{\rm n}^{\overline{\rm iso}}}{B_{\rm w}^{\rm iso}/B_{\rm w}^{\overline{\rm iso}}}=1,
\end{equation}
and assuming that the proportion of signal in the control regions ($\mathbb{B}$, $\mathbb{C}$ and $\mathbb{D}$)  is negligible compared to the proportion of background, the purity is derived in a data-driven approach (dd) as
\begin{equation}
\label{eq:ABCDpurity}
P_{\rm dd}=1-\frac {B_{\rm n}^{\overline{\rm iso}}/N_{\rm n}^{\rm iso}} {B_{\rm w}^{\overline{\rm iso}}/B_{\rm w}^{\rm iso}} = 1-\frac {N_{\rm n}^{\overline{\rm iso}}/N_{\rm n}^{\rm iso}} {N_{\rm w}^{\overline{\rm iso}}/N_{\rm w}^{\rm iso}}.
\end{equation}

Unfortunately,
both assumptions are valid only approximately, especially Eq.~\eqref{eq:ABCDproportionality}. In simulations with two jets in the final state that contribute only to background in all of the four zones, an evaluation of Eq.~\eqref{eq:ABCDproportionality} gives values of the order of 0.8 at $\ptg$~=~10\,\GeVc, increasing to about 1.7 for $\ptg > 40 $\,\GeVc, 
thus the ratio is in general different from unity.

In part, this is due to the fact that single photons from meson decays can have a higher value of  $p_{\rm T}^{\rm iso}$ than merged decay photons at the same \pt, because of the presence of the second photon 
from the meson decay in the isolation cone. 

Also, fluctuations in the cluster distributions, e.g.\ caused by overlapping showers from close particles originating from the same hard process, may lead to some energy contribution either to be included in the cluster candidate and increase its width, or not to be included and increase the isolation momentum, causing an anti-correlation of the two parameters.

Since these effects are purely due to particle kinematics and detector response, we use the simulation to estimate the bias this causes via
\begin{equation}
\label{eq:ABCDproportionalityMC}
\bigg ( \frac{B_{\rm n}^{\rm iso}/B_{\rm n}^{\overline{\rm iso}}}{B_{\rm w}^{\rm iso}/B_{\rm w}^{\overline{\rm iso}}} \bigg)_{\rm data}= \bigg ( \frac{B_{\rm n}^{\rm iso}/B_{\rm n}^{\overline{\rm iso}}}{B_{\rm w}^{\rm iso}/B_{\rm w}^{\overline{\rm iso}}} \bigg)_{\rm MC},
\end{equation}
where MC contains both jet-jet and $\gamma$-jet events scaled to their respective cross sections.  This implies replacing  Eq.~\eqref{eq:ABCDproportionality} by the relation given by Eq.~\eqref{eq:ABCDproportionalityMC} leading to the expression of the MC-corrected purity 

\begin{equation}
\label{eq:ABCDpurityMC}
P = 1-\bigg(\frac {N_{\rm n}^{\overline{\rm iso}}/N_{\rm n}^{\rm iso}} {N_{\rm w}^{\overline{\rm iso}}/N_{\rm w}^{\rm iso}}\bigg)_{\rm data} \times \bigg(\frac {B_{\rm n}^{\rm iso}/N_{\rm n}^{\overline{\rm iso}}} {N_{\rm w}^{\rm iso}/N_{\rm w}^{\overline{\rm iso}}}\bigg)_{\rm MC}  \equiv 1-\bigg(\frac {N_{\rm n}^{\overline{\rm iso}}/N_{\rm n}^{\rm iso}} {N_{\rm w}^{\overline{\rm iso}}/N_{\rm w}^{\rm iso}}\bigg)_{\rm data} \times \alpha_{\rm{MC}}.
\end{equation}

The difference between the degree of the correlation among isolation momentum and shower
shape distribution in data and in Monte-Carlo is another potential source of bias, as it
influences the validity of Eq.~\eqref{eq:ABCDproportionalityMC}. 
To check this, the dependence of the double ratio
\begin{equation}
\label{eq:doubleratio}
\frac {\left( N^{{\rm iso}}/N^{\overline{\rm iso}} \right) ^{\mathrm{data}}}
{\left( N^{{\rm iso}}/N^{\overline{\rm iso}} \right) ^{\mathrm{MC}}} = f\left( \shshlo \right)
\end{equation}
on the shower shape width \shshlo is studied in a region where the signal contribution is
expected to be negligible. If the correlation between the two variables is correctly 
reproduced  in the simulation, the double ratio is independent of $\shshlo$, i.e.\ it would be the same for wide and narrow clusters. The double ratio was found to be above unity, indicating a larger isolation probability in data than in simulations. This is mainly due to an imperfect  calibration of charged particle tracks which leads to some discrepancy between data and simulations in the estimate of the isolation energy from charged particle. 
However, since the correction introduced in Eq.~\eqref{eq:ABCDpurityMC} relies on a narrow-over-wide ratio, the overall normalisation in the double ratio of Eq.~\eqref{eq:doubleratio} does not enter the correction.

The double ratio was found to be consistent with a constant, but a possible  residual bias between data and MC has been estimated via
extrapolations by linear fits of the dependence on $\shshlo$ instead of
the assumption of a constant value. This consists of replacing the MC correction in Eq.~\eqref{eq:ABCDpurityMC} by a modified term:
\begin{equation}
\label{eq:CorrectionMCExtra}
\alpha_{{\rm MC}} \longmapsto \alpha_{{\rm MC}} \times \left(\frac{p_0+\sigma^2_\mathrm{long,n}.p_1}{p_0+\sigma^2_\mathrm{long,w}.p_1}\right),
\end{equation}

where $\sigma^2_\mathrm{long,n}$ and $\sigma^2_\mathrm{long,w}$ are the median values of the neutral cluster \shshlo\ distribution in the narrow and wide ranges, respectively, and $p_0$ and $p_1$ are the parameters of the linear fit of the double ratio $f(\shshlo)$.

These extrapolations have then been used in the estimate of the mean value and uncertainties of the purity. In this procedure, also a variation of the value of the isolation momentum required
for non-isolated clusters has been included for the Monte Carlo correction, in order to check the
variations due to the discrepancy of isolation fractions in data and in simulation (the overall normalisation).  Finally, 
the dependence of the results on the isolation momentum calculation is tested using 
only tracks when computing the isolation momentum 
($p_{\rm T}^{\rm iso}=\sum p_{\rm T}^{\rm track}$) and using an isolation criterion of $p_{\rm T}^{\rm iso} < 1.9$\,\GeVc\ and an anti-isolation cut of $p_{\rm T}^{\rm iso} > 2.9$\,\GeVc. These values have been chosen after comparing the isolated photon spectrum at generator level obtained using either neutral and charged particles or charged particles only in $\gamma$-jet simulations. The final purity is calculated as the mean value of all the results obtained from the different estimates varying:
\begin{itemize}
\item the isolation momentum definition (including or ignoring the contribution of $\sum p_{\rm T}^{\rm cluster}$),
\item the dependence of the isolation probability on the shower shape $\shshlo$ (extrapolation of $f(\shshlo)$),
\item the MC anti-isolation criterion (normalisation of $f(\shshlo)$).
\end{itemize}

\begin{figure}[ht]
\begin{center}
\includegraphics[width=0.5\textwidth]{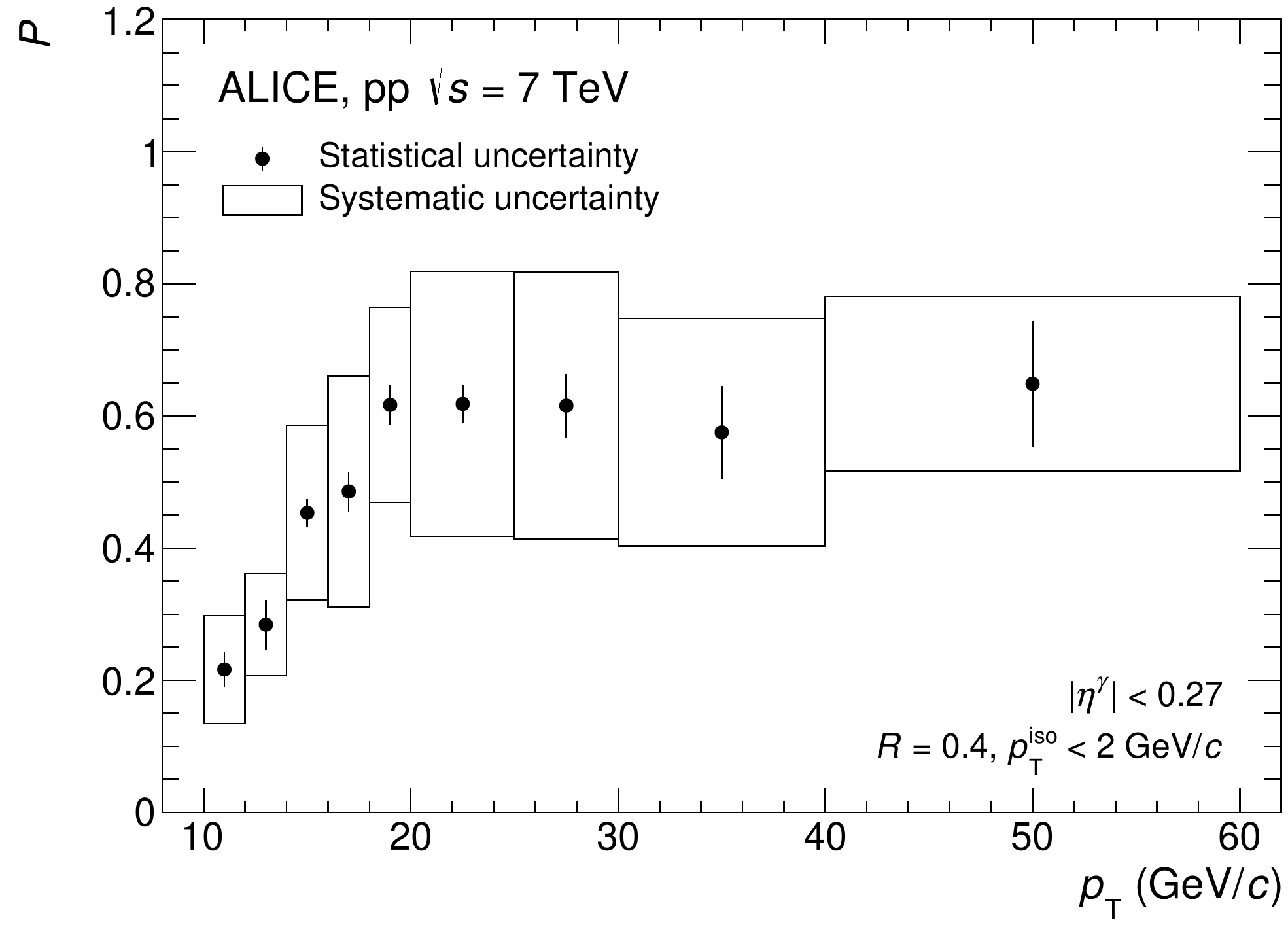} 
\end{center}
\caption{\label{fig:purityIsoPhoton} Isolated photon purity as a function of photon \pT.
}
\end{figure}

Fig.~\ref{fig:purityIsoPhoton} shows the purity calculated using Eq.~\eqref{eq:ABCDpurityMC} and averaged over the different approaches  listed above using also Eq.~\eqref{eq:CorrectionMCExtra}. 
The boxes indicate the systematic uncertainty whose estimation is explained in 
the next section. There is a large contamination at $\ptg = 10$\,\GeVc\ of 80\% 
that decreases and saturates at 40--50\% for $\ptg > 18$\,\GeVc. The contamination level defines the lower \ptg~that can be reached. 
Most of the contamination is due to \piz\ clusters (merged decay photons). 
Below 18\,\GeVc, this contamination is dominated by single (i.e.\ unmerged) decay photons from {\piz} mesons, 
the remaining contributors being mainly photons from $\eta$ mesons decay. Above 18\,\GeVc, a fraction of the merged \piz\ decay clusters have a narrow shower that satisfy the condition for the single photon signal, as illustrated in Fig.~\ref{fig:SigmaLongExample} (right). 
The $\ptg$ dependence of the purity is caused by an interplay of physics and 
detector effects. 
On one hand, the {\pt} spectra of prompt photons are harder 
than those of neutral pions, mainly because the latter undergo fragmentation, 
as also was found in pQCD calculations~\cite{Abeysekara:2010ze,Arleo:PhotonLHCYellow}.
For this reason, the $\gamma_{\rm dir} / \piz $ yield ratio rises with {\pt}, 
and the photon purity increases as well. 
Also, the probability to tag a photon as isolated varies with {\pt}. 
At higher \pt, isolation is less probable for a fixed isolation momentum. 
On the other hand, due to the decreasing opening angle at high \pt\ 
the contamination from \piz\ mesons increases with \pt. 
At $\pt = 20$\,\GeVc, 5\% of the \piz\ decay photons are found in the narrow 
shower shape region, and beyond {40\,\GeVc} this contribution rises to more than 25\%. 
The combined effect of these mechanisms leads to the rise of the purity at low \pt\ 
and a saturation for $\pt > 18$\,\GeVc.

The results obtained for the purity are comparable with those reported by CMS \cite{CMS2011,Khachatryan:2010fm} in the overlapping \pt{} range, whereas the purity obtained by ATLAS \cite{Aad:2010sp,Aad:2011tw} is significantly higher than our measurement due to the very high granularity of the first layers of its electromagnetic calorimeter, allowing a very good separation of single photon and \piz{} decay photon showers.

\subsection{Efficiency}

The photon reconstruction, identification and isolation  efficiency has been computed using PYTHIA 6 simulations of $\gamma$-jet processes
in which, for each event, a direct photon is emitted in the EMCal acceptance but, only 
those falling in the  fiducial acceptance are considered in the efficiency calculation. 

\begin{figure}[ht]
\begin{center}
\includegraphics[width=0.49\textwidth]{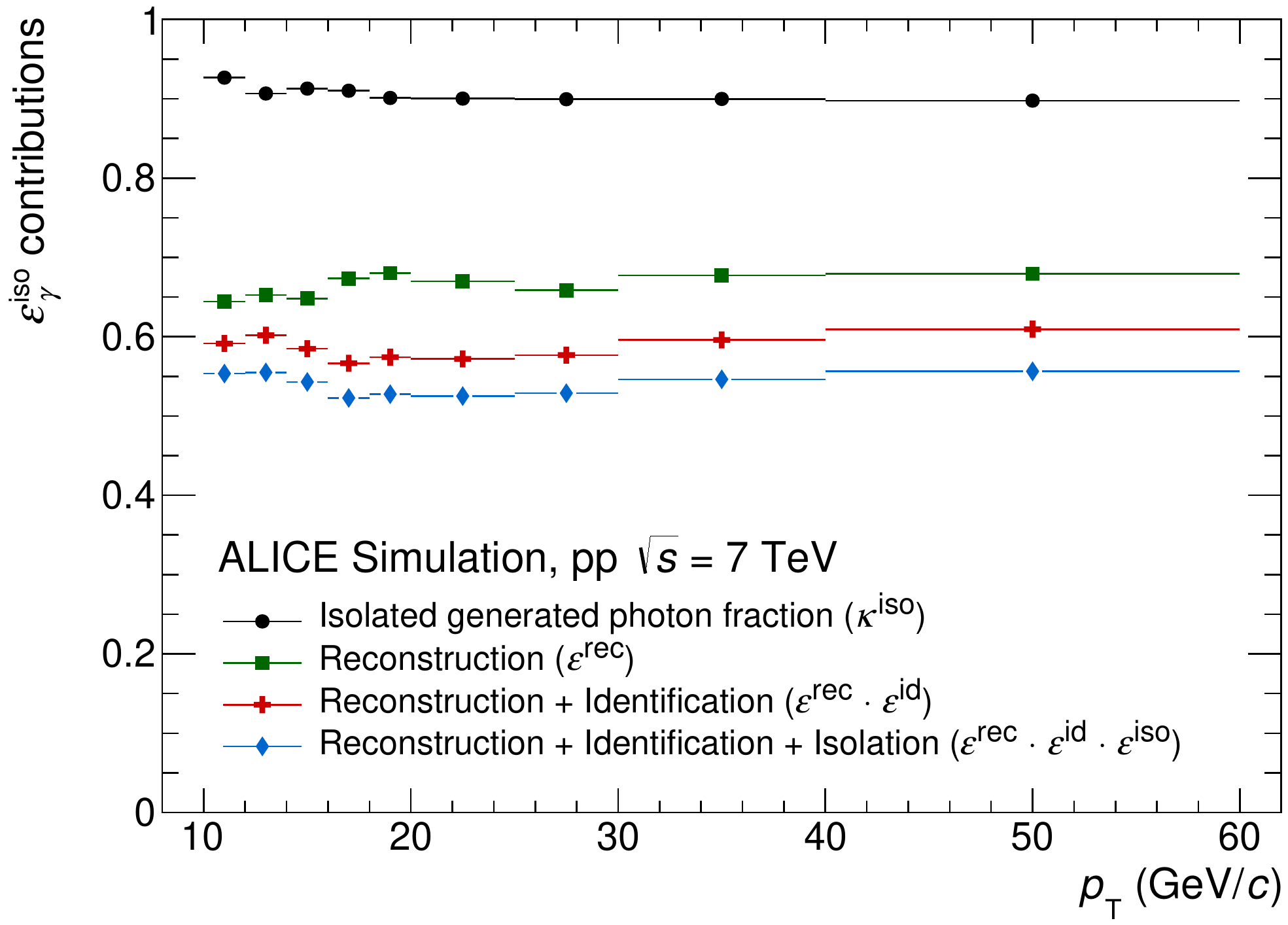}
\hfill
\includegraphics[width=0.49\textwidth]{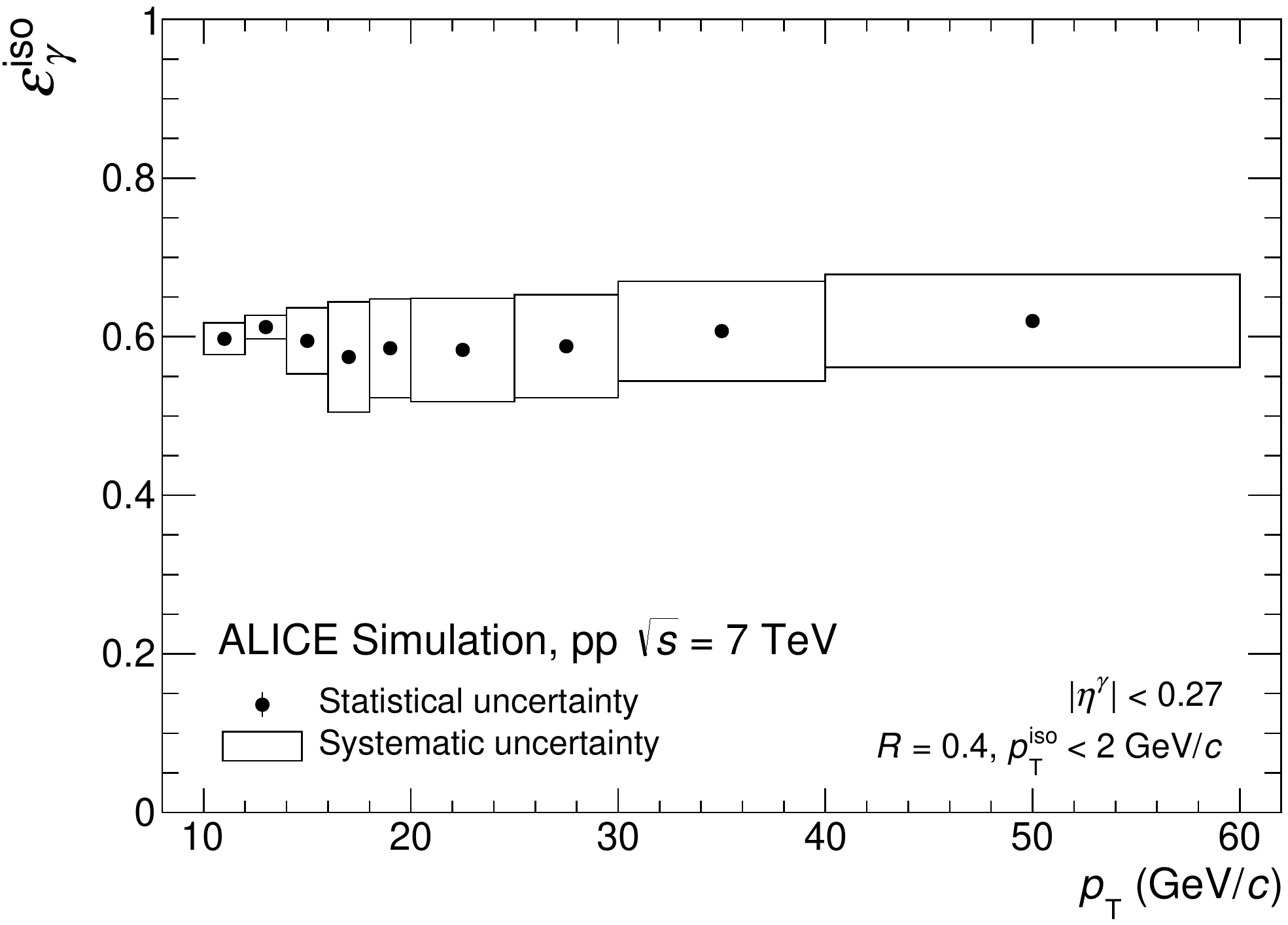}
\end{center}
\caption{\label{fig:EffIsoPhoton} (colour online) Different  contributions (reconstruction, identification, isolation) to the total efficiency (left) and total isolated photon efficiency (right) calculated using Eq.~\eqref{eq:efficiency}, all as a function of the reconstructed \pt. 
}
\end{figure}

The different analysis cuts contribute to the overall efficiency and the contributions are presented in the left panel of  Fig.~\ref{fig:EffIsoPhoton}. They are calculated as the ratio of spectra, where the denominator is the number of generated photons ${\rm d} N ^{\rm gen}_{\gamma} / {\rm d} p_{\rm T}^{\rm gen}$, and the numerators are the reconstructed spectra after different cuts, ${\rm d} N ^{\rm rec}_{\rm cut} / {\rm d} p_{\rm T}^{\rm rec}$:  
(i) the pure reconstruction efficiency of photons is $\epsilon^{\mathrm{rec}} \approx 70 \%$, (green squares), which includes losses due to excluded regions in the calorimeter and exclusion of clusters close to the border of EMCal supermodules, as well as bin migration due to energy resolution, 
(ii) applying in addition the photon identification reduces the efficiency by about 10\%, leading to $\epsilon^{\mathrm{rec}} \cdot \epsilon^{\mathrm{id}} \approx 60 \%$, (red crosses), which is mainly driven by the shower shape selection cuts used for the photon sample, 
(iii) combining it with the isolation criterion yields an efficiency $\epsilon^{\mathrm{rec}} \cdot \epsilon^{\mathrm{id}} \cdot \epsilon^{\mathrm{iso}} \approx 50 \%$, (blue diamonds).

In addition, the fraction $\kappa^{\mathrm{iso}}$ of generated photons which are isolated is represented by black filled circles in the figure. The total efficiency corresponds to the ratio of the reconstruction, identification and isolation efficiency as given in (iii) to the isolated generated photon fraction $\kappa^{\mathrm{iso}}$ and is then directly calculated as follows:
\begin{equation}
\label{eq:efficiency}
     \epsilon_{\gamma}^{\rm iso} = \frac{{\rm d} N ^{\rm rec}_{\rm n,\,iso}} { {\rm d} p_{\rm T}^{\rm rec}} {\Bigg /} \frac{{\rm d} N ^{\rm gen}_{\gamma,\,\rm iso}} { {\rm d} p_{\rm T}^{\rm gen}}
     \equiv \frac{\epsilon^{\mathrm{rec}} \epsilon^{\mathrm{id}} \epsilon^{\mathrm{iso}}}{\kappa^{\mathrm{iso}}},
\end{equation}
where $N^{\rm rec}_{\rm n,~iso} $ is the number of clusters which are reconstructed and identified as isolated photons and which are produced by a direct photon, and $N^{\rm gen}_{\gamma,~\rm iso}$ is the number of generated direct photons which pass the isolation cut in the same way as at the detector level.
The overall efficiency for the reconstruction of isolated photons is of the order of 60\% on average as shown in Fig.~\ref{fig:EffIsoPhoton} (right panel). To check the robustness of the efficiency calculation, the effect of a variation of the shape of the momentum spectrum in the Monte Carlo used has been studied and has been found to be negligible.

\subsection{Trigger efficiency and corrections}
The EMCal-L0 trigger efficiency $\epsilon_{\rm trig}$ is the probability that the trigger 
selects events when a photon is emitted in the EMCal acceptance. 
This analysis starts well above the trigger threshold (10~GeV compared to 5.5~GeV) where
the trigger efficiency is flat. 

The trigger efficiency is however not 100\%, because of two effects reducing the geometric coverage of the trigger compared to the EMCal acceptance: the sliding window technique can only be used within a single given TRU, and in addition, some TRU cards were inactive.
The efficiency is calculated from minimum bias events as the ratio of the number of events 
containing high energy clusters ($E>$~10~GeV, see cluster definition in Sect.~\ref{sec:clust}) and leaving a signal in the trigger over the total number of events with high energy clusters in the same sample.
It was estimated to be 
$\epsilon_{\rm trig}=$~0.90~$\pm$~0.06~(stat). The statistical uncertainty quoted here is completely correlated to that of the luminosity so that it will not be taken into account twice. Moreover, a 
bias in the trigger efficiency was found, which is due to  
synchronisation issues. Sometimes, the EMCal-L0 trigger selected events  
in the next bunch crossing, 50~ns after the nominal bunch crossing. 
The bias was estimated by calculating the ratio $\mathcal{C}$
of the number of clusters in a time window containing only the main bunch crossing 
over the number of clusters in a time window including the main and the next bunch crossing. 
The bias was found to be between 3\% and 8\% for trigger cluster \pt\ varying from 10 to 60~\GeVc{}, and the trigger efficiency is corrected for this effect. 
\section{Systematic uncertainties}
\label{sec:sys_unc}
\begin{figure}[h]
\begin{center}

\includegraphics[width=0.7\textwidth]{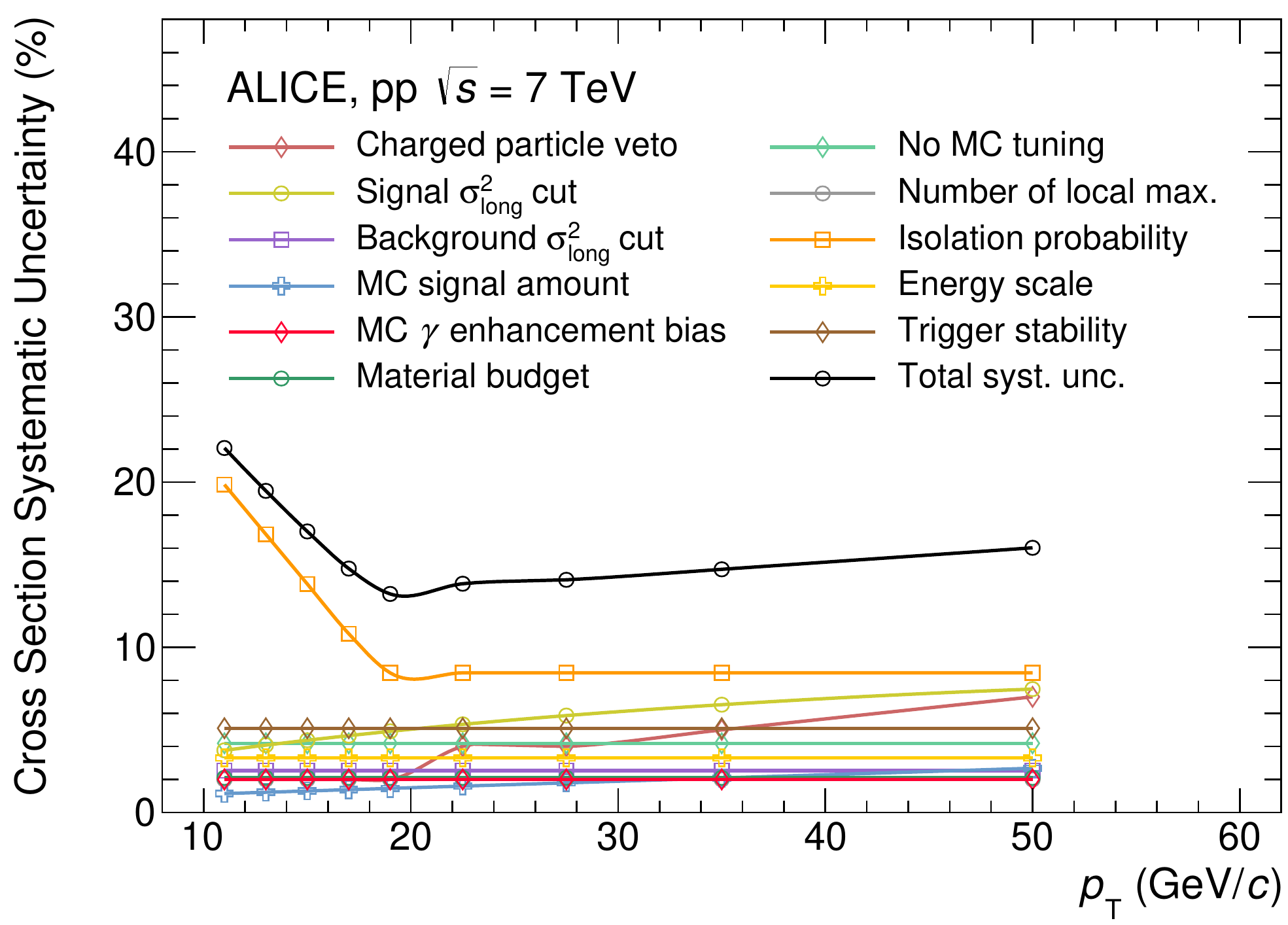}
\end{center}
\caption{\label{fig:systemacticIsoPhoton} (colour online) Contributions to the systematic uncertainty of the isolated photon cross section and their quadratic sum as a function of photon \pt.}
\end{figure}

Systematic uncertainties on the cross section measurement  are summarised in Table \ref{tab:systsummary} for two extreme transverse momentum bins used in the analysis and presented in Fig.~\ref{fig:systemacticIsoPhoton}. The uncertainties are treated as independent and thus summed in quadrature.
Though we present systematic uncertainties for intermediate quantities, like purity (Fig.~\ref{fig:purityIsoPhoton}) and efficiency (Fig.~\ref{fig:EffIsoPhoton}), they do not enter into the calculation of uncertainties of the final cross
section. Instead, systematic uncertainties of all sources are evaluated there directly.

\begin{table}[h]
    \centering
    \caption{Summary of uncorrelated relative systematic uncertainties in percent for selected $\ptg$ bins of the isolated photon measurement. The luminosity uncertainty of 9.5\% is not included in this table.}
    \label{tab:systsummary}
    \begin{tabular}{l l l }
        \hline
        $\ptg$ & 10--12\,\GeVc& 40--60\,\GeVc \\
        \hline
        Charged particle veto       & 2.0\%     & 7.0\% \\
        \shshlo{} signal range      & 3.7\%     & 7.5\%\\
        \shshlo{} background range  & 2.5\%     & 2.5\% \\
        MC signal amount            & 1.0\%    & 2.7\% \\
        MC $\gamma$ enhancement bias& 2.0\%     & 2.0\% \\
        No MC tuning                & 4.2\%     & 4.2\% \\
        Number of local maxima      & 2.0\%     & 2.0\% \\
        Isolation probability       & 20.0\%    & 8.5\%\\
        Energy scale                & 3.3\%     & 3.3 \%\\
        Trigger stability           & 5.1\%     & 5.1\% \\
        Material budget             & 2.1\%     & 2.1\% \\
        \hline
        Combined syst.\ unc.\ & 22.1\%    & 16.0\%\\
        Statistical unc.\     & 19.9\%    & 40.3\%\\
        \hline
    \end{tabular}
\end{table}

The uncertainties due to the choice of the photon cluster identification criteria in this analysis are evaluated via variations of cuts for the charged particle veto and the shower shape \shshlo\ for the photon selection.  

The uncertainty due to the charged particle veto was estimated by varying the parameters of the track \pt-dependent cuts for $\Delta \eta^{\rm residual}$ and $\Delta \varphi^{\rm residual} $. The resulting uncertainty on the cross section ranges from 2\% to 7\% from lower to larger $\ptg$.
The increase with \ptg\ is driven by the use of the charged particle veto in the cone activity measurement. For high values of \ptg, the in-cone activity is higher and the systematic uncertainty from the CPV is higher.

The choice of the signal range of the \shshlo\ of narrow photon-like showers, is
important for the efficiency calculation, but also influences the background estimate via a
“leakage” of photon showers to the control regions.
The uncertainty due to the choice of the signal range is estimated by varying the upper limit of the range and is found to lie between 3.7\% and 7.5\%, increasing with \ptg. Similarly, the uncertainty due to the choice of the background region (wide showers, i.e.\ large values of \shshlo) is investigated by moving the corresponding \shshlo{} interval to smaller and larger values. The estimated uncertainty is found to be 2.5\%. 

The Monte Carlo correction of the data-driven purity may also depend on the amount of signal in the simulation, mainly due to the aforementioned leakage effect. This is checked by changing the relative normalisation of the signal and background MC samples from the default value corresponding to the theoretical cross sections to a relative signal contribution that is 30\% larger or smaller than this default.The resulting uncertainty varies from 1\% for the lowest $\ptg$ to 2.7\% for the highest $\ptg$.  
The uncertainty related to the input particle bias produced by the event selection enhancing photons in simulation is 2\%. 

The description of the shower shape in simulations can also affect both the efficiency measurement and the MC correction of the purity. The associated uncertainty is found to be 4.2\%, 
estimated from the difference between standard simulations and those including modelling of the cross-talk observed in the EMCal readout cards. In addition, the sensitivity of the cross section to the number of local maxima of selected clusters is checked by accepting clusters with $N_{\rm LM}>2$. The resulting uncertainty amounts to 2\% in the cross section measurement.

Related to the purity, the bias due to the correlation between the two quantities used to estimate the contamination ($\sigma_{\mathrm{long}}^2$ and $p_{\mathrm{T}}^{\mathrm{iso}}$) is taken into account via a number of cut and method variations, which test the two main assumptions made in this estimate:
(i) the background isolation fraction is constant with respect to the shower shape $\sigma_{\mathrm{long}}^2$ and
(ii)  the isolation fraction is the same in data and simulation.
The different approaches used for these systematic checks are described in Sect.\ \ref{sec:purity}.
The systematic uncertainty assigned addressing these correlation effects is labelled as ``isolation probability'' and is obtained by the root mean square of the results obtained for the different checks mentioned above and in Sect.\ \ref{sec:purity}. The resulting uncertainty is estimated to vary from 20.\% ($\ptg = 10\,\GeVc$) to 8.5\% ($\ptg = 60\,\GeVc$). 

The decrease of uncertainties with \ptg\ is related to the increase of the purity of the photon candidates.

The uncertainty on the energy scale of the EMCal was estimated to be 0.8\%, determined from the analysis of test beam data \cite{ALLEN20106} and a comparison of the \piz\ mass-peak position and the energy-to-momentum ratio of electron tracks in data and Monte Carlo \cite{2017467}. This uncertainty amounts to 3.3\% in the cross section measurement.

The uncertainty on the trigger normalisation factor is 5.1\% and is estimated from the run-by-run variations of the number of reconstructed clusters with transverse momentum  above  $\ptg=$~10~\GeVc\ per event, corrected for the active detector area.

A material budget uncertainty accounting for the material of the different detectors traversed by photons before they reach the EMCal has been previously determined \cite{Acharya:2018dqe} and amounts to 2.1\% in the present measurement. The in-bunch pile-up uncertainty (affecting the raw yield via the isolation momentum) was found to be negligible, estimated by adding a random transverse momentum in the isolation cone   estimated from the in-cone energy of a random trigger in minimum bias events.

Fig.~\ref{fig:systemacticIsoPhoton} summarises the different sources of systematic uncertainties. The dominant source of uncertainty is the isolation probability, related mainly to the correlation of the two variables used for the purity estimation (\shshlo\ and $p_{\rm T}^{\rm iso}$), the discrepancy in isolation probability between data and MC and the definition of $p_{\rm T}^{\rm iso}$. The \ptg{} dependence of the total systematic uncertainty is expected and is related to the low purity at low \ptg.

\section{Results}
\label{sec:results}

The isolated direct photon production differential cross section can be obtained from the following equation:

\begin{equation}
  \frac{{\rm d}^{2} \sigma^{\gamma}}{{\rm d} \ptg ~{\rm d} \eta}  = \frac{1}{\mathcal{L} \epsilon_{\rm trig}\mathcal{C}}\frac{{\rm d}^{2} N^{\rm iso}_{n}} { {\rm d} \ptg ~{\rm d} \eta} \frac{P}{\epsilon_{\gamma}^{\rm iso}},
\end{equation}
where all the terms were described in the previous sections.

\begin{figure}[ht]
\begin{center}
\includegraphics[width=0.49\textwidth]{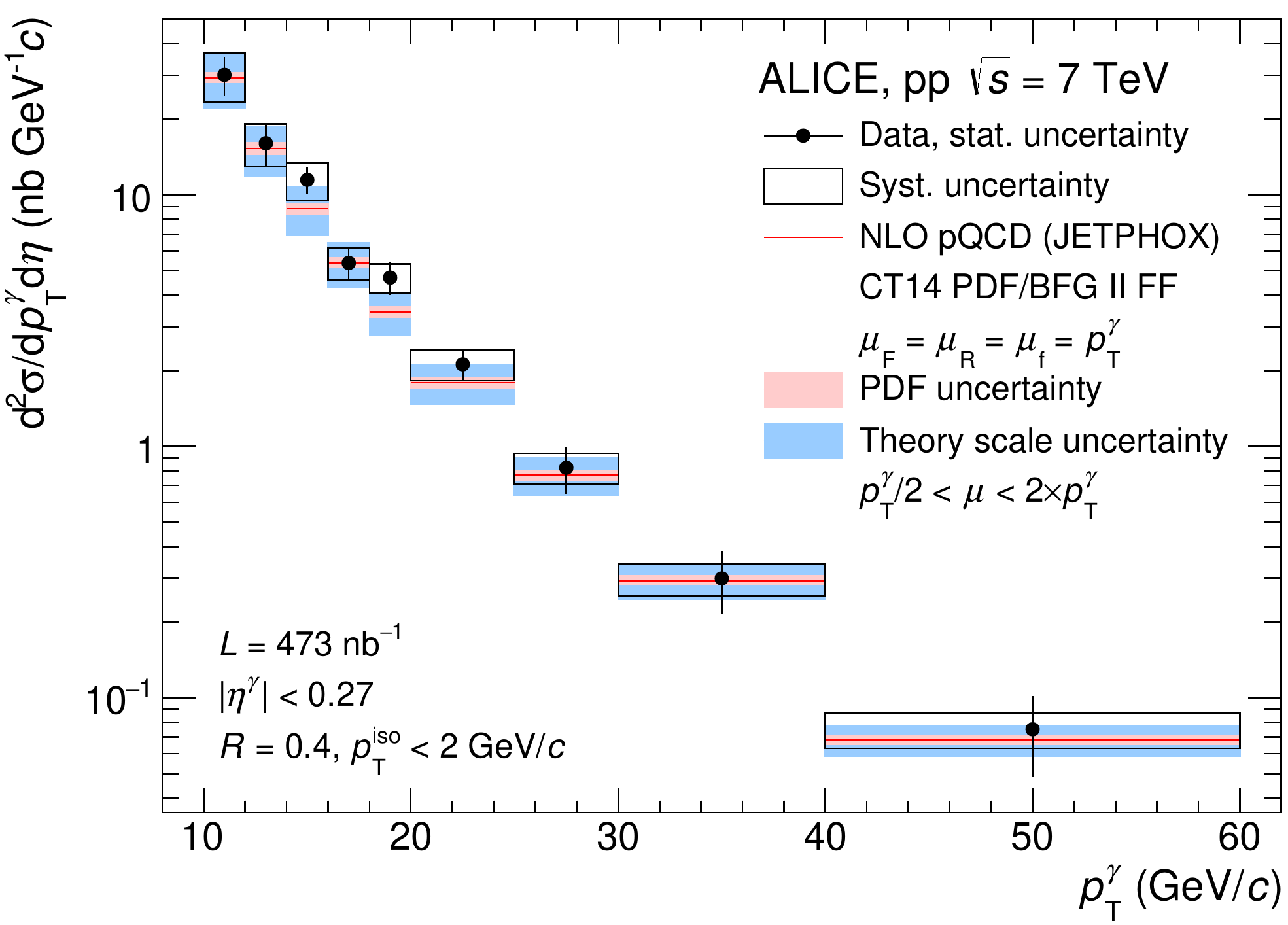} 
\hfill
\includegraphics[width=0.49\textwidth]{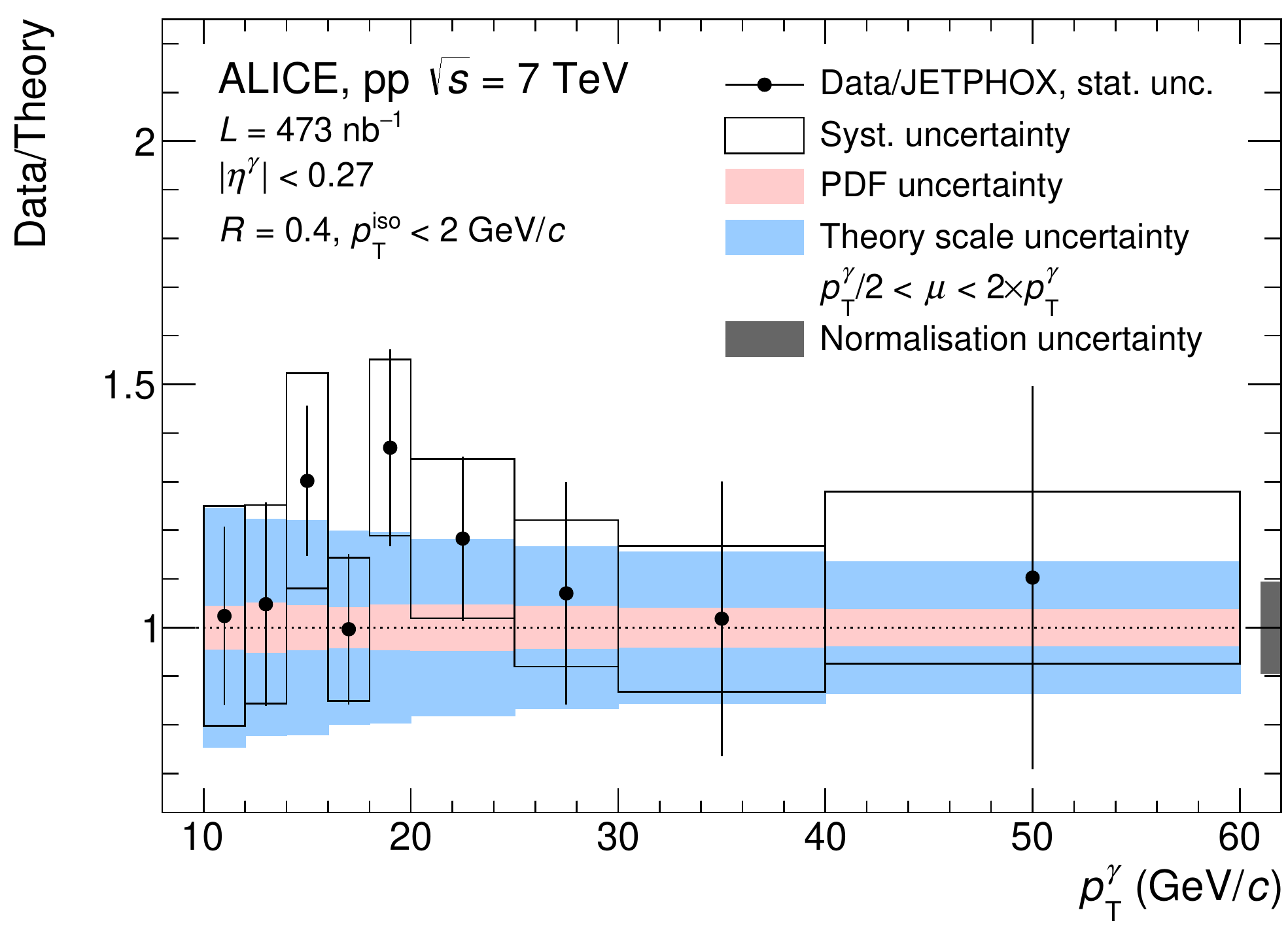}
\end{center}
\caption{\label{fig:isoPhotonCrossSection} (colour online)  Isolated photon differential cross section measured in pp collisions at $\sqrt s =7$~TeV (left plot). Error bars are statistical and boxes systematic uncertainties. The bands correspond to pQCD calculations with JETPHOX. The normalisation uncertainty explained in the text (9.5\%) is not included in the left panel and is presented as an overall box around unity in the right panel.} 
\end{figure}

Fig.~\ref{fig:isoPhotonCrossSection} shows the isolated photon cross section as a function of \ptg. Error bars indicate the statistical uncertainties and boxes the systematic uncertainties. An additional normalisation uncertainty of 9.5\%, which includes effects from the measurement of the total minimum bias cross section and effects due to the rejection factor from the EMCal triggering, is not displayed in the left panel of the figure.

The measurement is compared to next-to-leading order (NLO) pQCD calculations using JETPHOX 1.3.1~\cite{Fontannaz, Aurenche}. The parton distribution function (PDF) used is  CT14~\cite{CT14}, and the fragmentation function is BFG II~\cite{PFFth}. The central values of the predictions were obtained by choosing factorisation, normalisation and fragmentation scales equal to the photon transverse momentum ($\mu_{f}=\mu_{R}=\mu_{F}=\ptg$). Scale uncertainties were determined with a 7-point scale variation where $\mu_{R}$ and $\mu_{F}$ were varied by a factor of 2 up and down around $\ptg$, keeping the $\mu_{R}/\mu_{F}$ ratio between $1/2$ and 2. As uncertainties related to the PDF, the 56 eigenvector sets of CT14 were combined with the Hessian method~\cite{Butterworth:2015oua,Pumplin:2001ct}. 

The isolation criterion in the theory calculations corresponds to the hadronic energy at the partonic level within $R<0.4$ around the photon. The same threshold of $p_{\rm T}^{\rm iso} < $~2 \GeVc\ as in data is used.  The theoretical predictions are corrected to take into account the underlying event as well as the fragmentation in the isolation cone. This correction is estimated using $\gamma$-jet PYTHIA simulations as the fraction of generated photons which are isolated as shown in Fig~\ref{fig:EffIsoPhoton} (right). The theoretical predictions are computed in the same \ptg{} bins as for data.

Within uncertainties, the isolated photon cross section in data and theoretical predictions are in agreement for the full transverse momentum range measured as demonstrated by Fig.~\ref{fig:isoPhotonCrossSection} (right). 

\begin{figure}[t]
    \centering
  
    \includegraphics[width=0.7\textwidth]{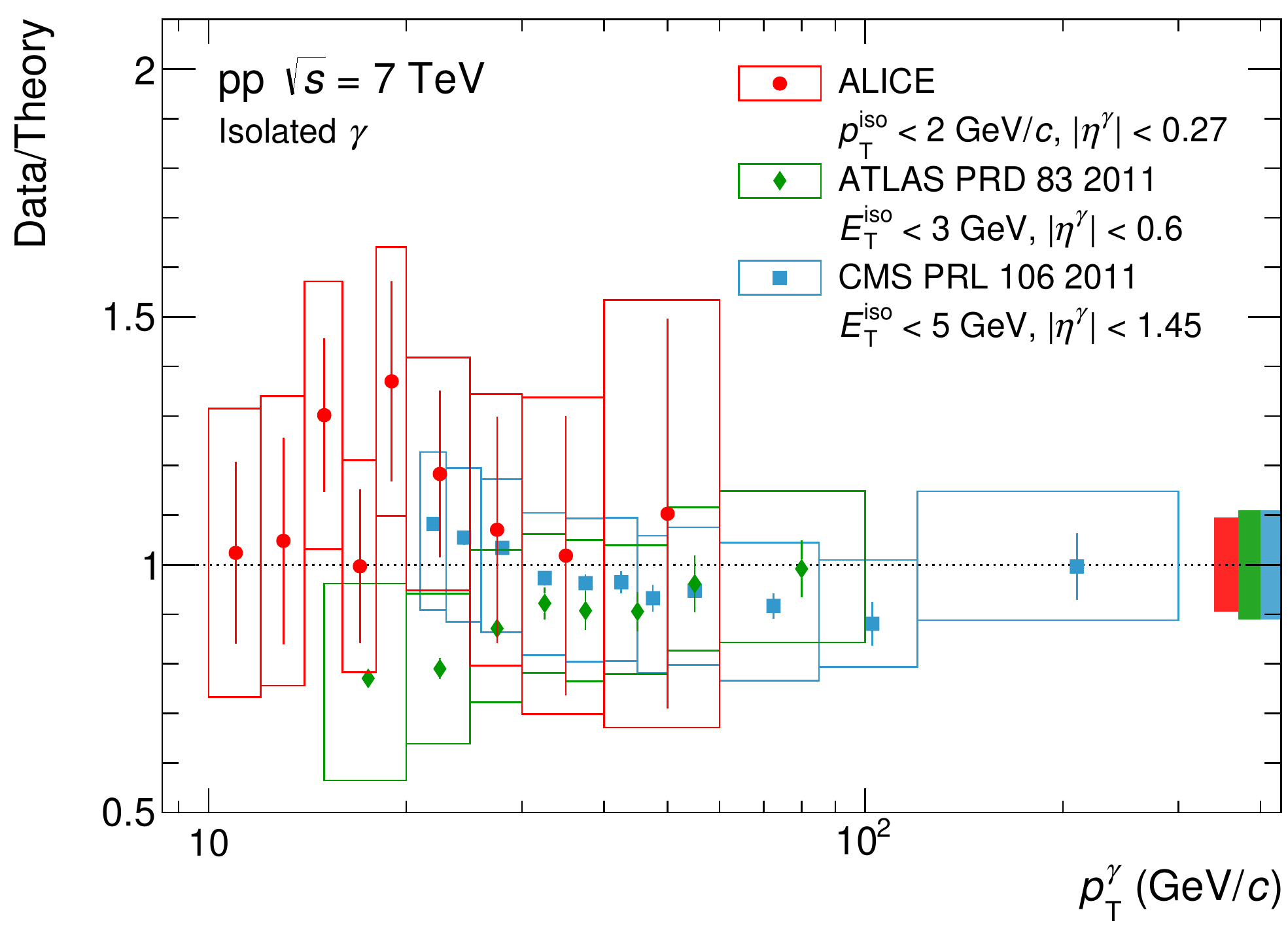} 
    \caption{\label{fig:isoPhotonCrossSectionComp}(color online) Ratio between differential cross section measurements and theory predictions for ATLAS~\cite{Aad:2010sp}, CMS~\cite{Khachatryan:2010fm} and ALICE. Theory predictions are obtained with JETPHOX and CTEQ 6.6 PDFs \cite{CTEQ66} for ATLAS, and JETPHOX and CT14 PDFs \cite{CT14} for CMS and ALICE. Only experimental uncertainties are shown here. Error bars are statistical and boxes are the quadratic sum of statistical and systematic uncertainties. The normalisation uncertainty of each experiment is presented as an overall box around unity.}
\end{figure}

Fig.~\ref{fig:isoPhotonCrossSectionComp} compares the ratios of measured differential isolated photon cross sections to theoretical predictions from three  different LHC experiments, namely ALICE, ATLAS \cite{Aad:2010sp} and CMS \cite{Khachatryan:2010fm}. 
The comparison is done on ratios of data to similar predictions since the isolation criteria differ among these experiments such that a direct comparison of the isolated photon cross sections is not fully adequate. The ATLAS and CMS experiments use larger values for $p_{\rm T}^{\rm iso}$. In JETPHOX predictions, increasing the isolation  threshold should reflect in a larger fragmentation contribution in the total cross section without necessarily increasing the total isolated photon cross section compared to smaller isolation criteria. However, the data to theory ratios should be consistent between the experiments as it is observed in Fig.~\ref{fig:isoPhotonCrossSectionComp}.

The ALICE measurement extends the \ptg ~range to lower values than ATLAS, which has measured the isolated photon cross section at mid-rapidity for $\ptg \ge 15$~\GeVc{} in the same collision system. All experiments agree with pQCD predictions within theoretical and experimental uncertainties.

\begin{figure}[htbp]
\begin{center}
\includegraphics[width=0.7\textwidth]{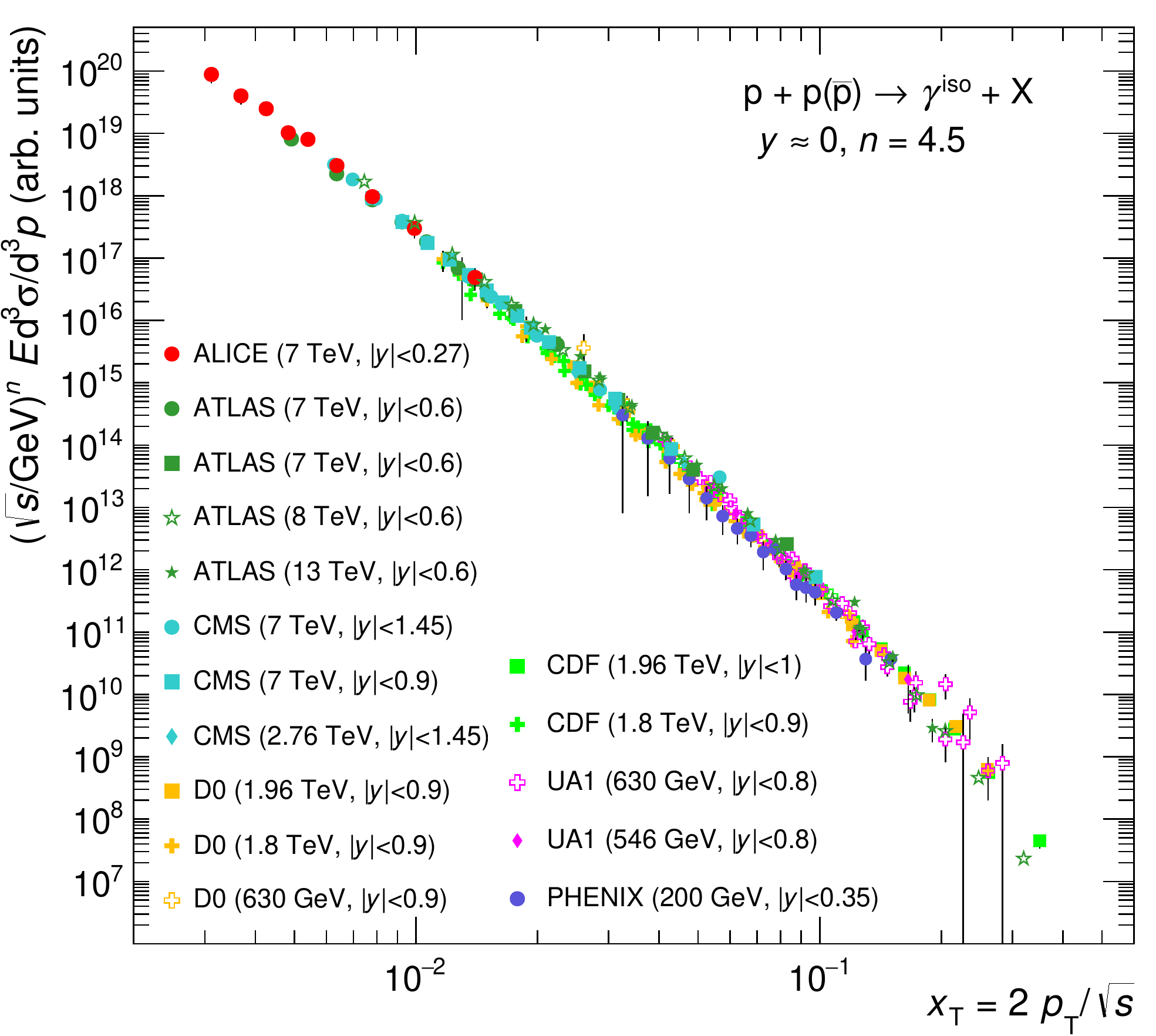} 

\end{center}
\caption{\label{fig:isoPhotonWorld}(colour online) ALICE data compared to the world's data of isolated photon spectra measured in pp and p$\overline{\rm p}$ collisions as a function of $x_{\rm T }$ where the invariant cross sections have been scaled by $(\sqrt{s})^n$ with $n=4.5$ compiled in Ref.~\cite{DENTERRIA2012311}. For this comparison only the results covering mid-rapidity are shown.}
\end{figure}

For a comparison of cross sections measured at different $\sqrt{s}$, it is more appropriate to use the variable $x_{\mathrm{T}}$ as defined in Eq.~\eqref{eq-bjorkenx}, which is also closely related to Bjorken $x$ \cite{Bjorken}.
A compilation of all available data on isolated photon cross section measurements in collider experiments has been performed in \cite{DENTERRIA2012311} and  all $x_{\rm T }$ spectra were compatible with a single curve when scaled by $(\sqrt{s})^n$ with $n=4.5$.
The ALICE measurement is compared to those data including also latest LHC measurements and the result is presented on Fig.~\ref{fig:isoPhotonWorld}. 
The ALICE measurement, as anticipated, allows us to  extend  the $x_{\rm T }$ reach to lower values, and is in agreement with the $n=4.5$ scaling, suggesting that all data are sensitive to the same production mechanisms. 
However, the value $n=4.5$ deviates from the $1/(\pt)^{n=4}$ dependence expected for the leading-twist partonic production mechanisms. 
This may be due to effects like the running coupling and the evolution of PDFs, but could also indicate significant contributions from fragmentation photons and higher twist diagrams \cite{PhysRevLett.105.062002}.

\section{Conclusions}
\label{sec:conclusion}
The isolated photon differential cross section in pp collisions at $\sqrt s = 7$~TeV is measured by the ALICE experiment at mid-rapidity in the transverse momentum range from 10 to 60~\GeVc.
Results are compared to ATLAS and CMS results and to pQCD calculations. The mutual agreement of the data sets with theory supports the theoretical calculations and demonstrates the consistency of the different measurements.

The current measurement extends the lower limit of {\ptg} to a smaller value compared to previous measurements by other experiments. This capability of ALICE will be useful for future studies of isolated photon cross sections and correlations of isolated photons to jets or hadrons in high-statistics data samples, in particular also for studying medium-induced modifications of hard probes. The measurement also opens up the possibility to access lower Bjorken-$x$. While in pp collisions this measurement, in spite of its lower {\pt} reach, may not provide strong constraints on the low-$x$ PDFs, this should be much more promising in nuclear collisions due to the larger uncertainties of nuclear PDFs.

\FloatBarrier

\newenvironment{acknowledgement}{\relax}{\relax}
\begin{acknowledgement}
\section*{Acknowledgements}
The authors would like to thank D.\ d'Enterria for fruitful discussions and the CMS Collaboration for providing details on previous measurements.

The ALICE Collaboration would like to thank all its engineers and technicians for their invaluable contributions to the construction of the experiment and the CERN accelerator teams for the outstanding performance of the LHC complex.
The ALICE Collaboration gratefully acknowledges the resources and support provided by all Grid centres and the Worldwide LHC Computing Grid (WLCG) collaboration.
The ALICE Collaboration acknowledges the following funding agencies for their support in building and running the ALICE detector:
A. I. Alikhanyan National Science Laboratory (Yerevan Physics Institute) Foundation (ANSL), State Committee of Science and World Federation of Scientists (WFS), Armenia;
Austrian Academy of Sciences, Austrian Science Fund (FWF): [M 2467-N36] and Nationalstiftung f\"{u}r Forschung, Technologie und Entwicklung, Austria;
Ministry of Communications and High Technologies, National Nuclear Research Center, Azerbaijan;
Conselho Nacional de Desenvolvimento Cient\'{\i}fico e Tecnol\'{o}gico (CNPq), Universidade Federal do Rio Grande do Sul (UFRGS), Financiadora de Estudos e Projetos (Finep) and Funda\c{c}\~{a}o de Amparo \`{a} Pesquisa do Estado de S\~{a}o Paulo (FAPESP), Brazil;
Ministry of Science \& Technology of China (MSTC), National Natural Science Foundation of China (NSFC) and Ministry of Education of China (MOEC) , China;
Croatian Science Foundation and Ministry of Science and Education, Croatia;
Centro de Aplicaciones Tecnol\'{o}gicas y Desarrollo Nuclear (CEADEN), Cubaenerg\'{\i}a, Cuba;
Ministry of Education, Youth and Sports of the Czech Republic, Czech Republic;
The Danish Council for Independent Research | Natural Sciences, the Carlsberg Foundation and Danish National Research Foundation (DNRF), Denmark;
Helsinki Institute of Physics (HIP), Finland;
Commissariat \`{a} l'Energie Atomique (CEA), Institut National de Physique Nucl\'{e}aire et de Physique des Particules (IN2P3) and Centre National de la Recherche Scientifique (CNRS) and R\'{e}gion des  Pays de la Loire, France;
Bundesministerium f\"{u}r Bildung und Forschung (BMBF) and GSI Helmholtzzentrum f\"{u}r Schwerionenforschung GmbH, Germany;
General Secretariat for Research and Technology, Ministry of Education, Research and Religions, Greece;
National Research, Development and Innovation Office, Hungary;
Department of Atomic Energy Government of India (DAE), Department of Science and Technology, Government of India (DST), University Grants Commission, Government of India (UGC) and Council of Scientific and Industrial Research (CSIR), India;
Indonesian Institute of Science, Indonesia;
Centro Fermi - Museo Storico della Fisica e Centro Studi e Ricerche Enrico Fermi and Istituto Nazionale di Fisica Nucleare (INFN), Italy;
Institute for Innovative Science and Technology , Nagasaki Institute of Applied Science (IIST), Japan Society for the Promotion of Science (JSPS) KAKENHI and Japanese Ministry of Education, Culture, Sports, Science and Technology (MEXT), Japan;
Consejo Nacional de Ciencia (CONACYT) y Tecnolog\'{i}a, through Fondo de Cooperaci\'{o}n Internacional en Ciencia y Tecnolog\'{i}a (FONCICYT) and Direcci\'{o}n General de Asuntos del Personal Academico (DGAPA), Mexico;
Nederlandse Organisatie voor Wetenschappelijk Onderzoek (NWO), Netherlands;
The Research Council of Norway, Norway;
Commission on Science and Technology for Sustainable Development in the South (COMSATS), Pakistan;
Pontificia Universidad Cat\'{o}lica del Per\'{u}, Peru;
Ministry of Science and Higher Education and National Science Centre, Poland;
Korea Institute of Science and Technology Information and National Research Foundation of Korea (NRF), Republic of Korea;
Ministry of Education and Scientific Research, Institute of Atomic Physics and Ministry of Research and Innovation and Institute of Atomic Physics, Romania;
Joint Institute for Nuclear Research (JINR), Ministry of Education and Science of the Russian Federation, National Research Centre Kurchatov Institute, Russian Science Foundation and Russian Foundation for Basic Research, Russia;
Ministry of Education, Science, Research and Sport of the Slovak Republic, Slovakia;
National Research Foundation of South Africa, South Africa;
Swedish Research Council (VR) and Knut \& Alice Wallenberg Foundation (KAW), Sweden;
European Organization for Nuclear Research, Switzerland;
National Science and Technology Development Agency (NSDTA), Suranaree University of Technology (SUT) and Office of the Higher Education Commission under NRU project of Thailand, Thailand;
Turkish Atomic Energy Agency (TAEK), Turkey;
National Academy of  Sciences of Ukraine, Ukraine;
Science and Technology Facilities Council (STFC), United Kingdom;
National Science Foundation of the United States of America (NSF) and United States Department of Energy, Office of Nuclear Physics (DOE NP), United States of America.    
\end{acknowledgement}


\bibliographystyle{utphys}   
\bibliography{biblio}
\newpage
\appendix
\section{The ALICE Collaboration}
\label{app:collab}

\begingroup
\small
\begin{flushleft}
S.~Acharya\Irefn{org141}\And 
D.~Adamov\'{a}\Irefn{org93}\And 
S.P.~Adhya\Irefn{org141}\And 
A.~Adler\Irefn{org73}\And 
J.~Adolfsson\Irefn{org79}\And 
M.M.~Aggarwal\Irefn{org98}\And 
G.~Aglieri Rinella\Irefn{org34}\And 
M.~Agnello\Irefn{org31}\And 
N.~Agrawal\Irefn{org10}\textsuperscript{,}\Irefn{org48}\textsuperscript{,}\Irefn{org53}\And 
Z.~Ahammed\Irefn{org141}
% ...
\And
\renewcommand\labelenumi{\textsuperscript{\theenumi}~}

\section*{Affiliation notes}
\renewcommand\theenumi{\roman{enumi}}
\begin{Authlist}
\item \Adef{org*}Deceased
\item \Adef{orgI}Dipartimento DET del Politecnico di Torino, Turin, Italy
\item \Adef{orgII}M.V. Lomonosov Moscow State University, D.V. Skobeltsyn Institute of Nuclear, Physics, Moscow, Russia
\item \Adef{orgIII}Department of Applied Physics, Aligarh Muslim University, Aligarh, India
\item \Adef{orgIV}Institute of Theoretical Physics, University of Wroclaw, Poland
\end{Authlist}

\section*{Collaboration Institutes}
\renewcommand\theenumi{\arabic{enumi}~}
\begin{Authlist}
\item \Idef{org1}A.I. Alikhanyan National Science Laboratory (Yerevan Physics Institute) Foundation, Yerevan, Armenia
\item \Idef{org2}Bogolyubov Institute for Theoretical Physics, National Academy of Sciences of Ukraine, Kiev, Ukraine
\item \Idef{org3}Bose Institute, Department of Physics  and Centre for Astroparticle Physics and Space Science (CAPSS), Kolkata, India
\item \Idef{org4}Budker Institute for Nuclear Physics, Novosibirsk, Russia
\item \Idef{org5}California Polytechnic State University, San Luis Obispo, California, United States
\item \Idef{org6}Central China Normal University, Wuhan, China
\item \Idef{org7}Centre de Calcul de l'IN2P3, Villeurbanne, Lyon, France
\item \Idef{org8}Centro de Aplicaciones Tecnol\'{o}gicas y Desarrollo Nuclear (CEADEN), Havana, Cuba
\item \Idef{org9}Centro de Investigaci\'{o}n y de Estudios Avanzados (CINVESTAV), Mexico City and M\'{e}rida, Mexico
\item \Idef{org10}Centro Fermi - Museo Storico della Fisica e Centro Studi e Ricerche ``Enrico Fermi', Rome, Italy
\item \Idef{org11}Chicago State University, Chicago, Illinois, United States
\item \Idef{org12}China Institute of Atomic Energy, Beijing, China
\item \Idef{org13}Chonbuk National University, Jeonju, Republic of Korea
\item \Idef{org14}Comenius University Bratislava, Faculty of Mathematics, Physics and Informatics, Bratislava, Slovakia
\item \Idef{org15}COMSATS University Islamabad, Islamabad, Pakistan
\item \Idef{org16}Creighton University, Omaha, Nebraska, United States
\item \Idef{org17}Department of Physics, Aligarh Muslim University, Aligarh, India
\item \Idef{org18}Department of Physics, Pusan National University, Pusan, Republic of Korea
\item \Idef{org19}Department of Physics, Sejong University, Seoul, Republic of Korea
\item \Idef{org20}Department of Physics, University of California, Berkeley, California, United States
\item \Idef{org21}Department of Physics, University of Oslo, Oslo, Norway
\item \Idef{org22}Department of Physics and Technology, University of Bergen, Bergen, Norway
\item \Idef{org23}Dipartimento di Fisica dell'Universit\`{a} 'La Sapienza' and Sezione INFN, Rome, Italy
\item \Idef{org24}Dipartimento di Fisica dell'Universit\`{a} and Sezione INFN, Cagliari, Italy
\item \Idef{org25}Dipartimento di Fisica dell'Universit\`{a} and Sezione INFN, Trieste, Italy
\item \Idef{org26}Dipartimento di Fisica dell'Universit\`{a} and Sezione INFN, Turin, Italy
\item \Idef{org27}Dipartimento di Fisica e Astronomia dell'Universit\`{a} and Sezione INFN, Bologna, Italy
\item \Idef{org28}Dipartimento di Fisica e Astronomia dell'Universit\`{a} and Sezione INFN, Catania, Italy
\item \Idef{org29}Dipartimento di Fisica e Astronomia dell'Universit\`{a} and Sezione INFN, Padova, Italy
\item \Idef{org30}Dipartimento di Fisica `E.R.~Caianiello' dell'Universit\`{a} and Gruppo Collegato INFN, Salerno, Italy
\item \Idef{org31}Dipartimento DISAT del Politecnico and Sezione INFN, Turin, Italy
\item \Idef{org32}Dipartimento di Scienze e Innovazione Tecnologica dell'Universit\`{a} del Piemonte Orientale and INFN Sezione di Torino, Alessandria, Italy
\item \Idef{org33}Dipartimento Interateneo di Fisica `M.~Merlin' and Sezione INFN, Bari, Italy
\item \Idef{org34}European Organization for Nuclear Research (CERN), Geneva, Switzerland
\item \Idef{org35}Faculty of Electrical Engineering, Mechanical Engineering and Naval Architecture, University of Split, Split, Croatia
\item \Idef{org36}Faculty of Engineering and Science, Western Norway University of Applied Sciences, Bergen, Norway
\item \Idef{org37}Faculty of Nuclear Sciences and Physical Engineering, Czech Technical University in Prague, Prague, Czech Republic
\item \Idef{org38}Faculty of Science, P.J.~\v{S}af\'{a}rik University, Ko\v{s}ice, Slovakia
\item \Idef{org39}Frankfurt Institute for Advanced Studies, Johann Wolfgang Goethe-Universit\"{a}t Frankfurt, Frankfurt, Germany
\item \Idef{org40}Gangneung-Wonju National University, Gangneung, Republic of Korea
\item \Idef{org41}Gauhati University, Department of Physics, Guwahati, India
\item \Idef{org42}Helmholtz-Institut f\"{u}r Strahlen- und Kernphysik, Rheinische Friedrich-Wilhelms-Universit\"{a}t Bonn, Bonn, Germany
\item \Idef{org43}Helsinki Institute of Physics (HIP), Helsinki, Finland
\item \Idef{org44}High Energy Physics Group,  Universidad Aut\'{o}noma de Puebla, Puebla, Mexico
\item \Idef{org45}Hiroshima University, Hiroshima, Japan
\item \Idef{org46}Hochschule Worms, Zentrum  f\"{u}r Technologietransfer und Telekommunikation (ZTT), Worms, Germany
\item \Idef{org47}Horia Hulubei National Institute of Physics and Nuclear Engineering, Bucharest, Romania
\item \Idef{org48}Indian Institute of Technology Bombay (IIT), Mumbai, India
\item \Idef{org49}Indian Institute of Technology Indore, Indore, India
\item \Idef{org50}Indonesian Institute of Sciences, Jakarta, Indonesia
\item \Idef{org51}INFN, Laboratori Nazionali di Frascati, Frascati, Italy
\item \Idef{org52}INFN, Sezione di Bari, Bari, Italy
\item \Idef{org53}INFN, Sezione di Bologna, Bologna, Italy
\item \Idef{org54}INFN, Sezione di Cagliari, Cagliari, Italy
\item \Idef{org55}INFN, Sezione di Catania, Catania, Italy
\item \Idef{org56}INFN, Sezione di Padova, Padova, Italy
\item \Idef{org57}INFN, Sezione di Roma, Rome, Italy
\item \Idef{org58}INFN, Sezione di Torino, Turin, Italy
\item \Idef{org59}INFN, Sezione di Trieste, Trieste, Italy
\item \Idef{org60}Inha University, Incheon, Republic of Korea
\item \Idef{org61}Institut de Physique Nucl\'{e}aire d'Orsay (IPNO), Institut National de Physique Nucl\'{e}aire et de Physique des Particules (IN2P3/CNRS), Universit\'{e} de Paris-Sud, Universit\'{e} Paris-Saclay, Orsay, France
\item \Idef{org62}Institute for Nuclear Research, Academy of Sciences, Moscow, Russia
\item \Idef{org63}Institute for Subatomic Physics, Utrecht University/Nikhef, Utrecht, Netherlands
\item \Idef{org64}Institute of Experimental Physics, Slovak Academy of Sciences, Ko\v{s}ice, Slovakia
\item \Idef{org65}Institute of Physics, Homi Bhabha National Institute, Bhubaneswar, India
\item \Idef{org66}Institute of Physics of the Czech Academy of Sciences, Prague, Czech Republic
\item \Idef{org67}Institute of Space Science (ISS), Bucharest, Romania
\item \Idef{org68}Institut f\"{u}r Kernphysik, Johann Wolfgang Goethe-Universit\"{a}t Frankfurt, Frankfurt, Germany
\item \Idef{org69}Instituto de Ciencias Nucleares, Universidad Nacional Aut\'{o}noma de M\'{e}xico, Mexico City, Mexico
\item \Idef{org70}Instituto de F\'{i}sica, Universidade Federal do Rio Grande do Sul (UFRGS), Porto Alegre, Brazil
\item \Idef{org71}Instituto de F\'{\i}sica, Universidad Nacional Aut\'{o}noma de M\'{e}xico, Mexico City, Mexico
\item \Idef{org72}iThemba LABS, National Research Foundation, Somerset West, South Africa
\item \Idef{org73}Johann-Wolfgang-Goethe Universit\"{a}t Frankfurt Institut f\"{u}r Informatik, Fachbereich Informatik und Mathematik, Frankfurt, Germany
\item \Idef{org74}Joint Institute for Nuclear Research (JINR), Dubna, Russia
\item \Idef{org75}Korea Institute of Science and Technology Information, Daejeon, Republic of Korea
\item \Idef{org76}KTO Karatay University, Konya, Turkey
\item \Idef{org77}Laboratoire de Physique Subatomique et de Cosmologie, Universit\'{e} Grenoble-Alpes, CNRS-IN2P3, Grenoble, France
\item \Idef{org78}Lawrence Berkeley National Laboratory, Berkeley, California, United States
\item \Idef{org79}Lund University Department of Physics, Division of Particle Physics, Lund, Sweden
\item \Idef{org80}Nagasaki Institute of Applied Science, Nagasaki, Japan
\item \Idef{org81}Nara Women{'}s University (NWU), Nara, Japan
\item \Idef{org82}National and Kapodistrian University of Athens, School of Science, Department of Physics , Athens, Greece
\item \Idef{org83}National Centre for Nuclear Research, Warsaw, Poland
\item \Idef{org84}National Institute of Science Education and Research, Homi Bhabha National Institute, Jatni, India
\item \Idef{org85}National Nuclear Research Center, Baku, Azerbaijan
\item \Idef{org86}National Research Centre Kurchatov Institute, Moscow, Russia
\item \Idef{org87}Niels Bohr Institute, University of Copenhagen, Copenhagen, Denmark
\item \Idef{org88}Nikhef, National institute for subatomic physics, Amsterdam, Netherlands
\item \Idef{org89}NRC Kurchatov Institute IHEP, Protvino, Russia
\item \Idef{org90}NRC << Kurchatov Institut >> - ITEP, Moscow, Russia
\item \Idef{org91}NRNU Moscow Engineering Physics Institute, Moscow, Russia
\item \Idef{org92}Nuclear Physics Group, STFC Daresbury Laboratory, Daresbury, United Kingdom
\item \Idef{org93}Nuclear Physics Institute of the Czech Academy of Sciences, \v{R}e\v{z} u Prahy, Czech Republic
\item \Idef{org94}Oak Ridge National Laboratory, Oak Ridge, Tennessee, United States
\item \Idef{org95}Ohio State University, Columbus, Ohio, United States
\item \Idef{org96}Petersburg Nuclear Physics Institute, Gatchina, Russia
\item \Idef{org97}Physics department, Faculty of science, University of Zagreb, Zagreb, Croatia
\item \Idef{org98}Physics Department, Panjab University, Chandigarh, India
\item \Idef{org99}Physics Department, University of Jammu, Jammu, India
\item \Idef{org100}Physics Department, University of Rajasthan, Jaipur, India
\item \Idef{org101}Physikalisches Institut, Eberhard-Karls-Universit\"{a}t T\"{u}bingen, T\"{u}bingen, Germany
\item \Idef{org102}Physikalisches Institut, Ruprecht-Karls-Universit\"{a}t Heidelberg, Heidelberg, Germany
\item \Idef{org103}Physik Department, Technische Universit\"{a}t M\"{u}nchen, Munich, Germany
\item \Idef{org104}Politecnico di Bari, Bari, Italy
\item \Idef{org105}Research Division and ExtreMe Matter Institute EMMI, GSI Helmholtzzentrum f\"ur Schwerionenforschung GmbH, Darmstadt, Germany
\item \Idef{org106}Rudjer Bo\v{s}kovi\'{c} Institute, Zagreb, Croatia
\item \Idef{org107}Russian Federal Nuclear Center (VNIIEF), Sarov, Russia
\item \Idef{org108}Saha Institute of Nuclear Physics, Homi Bhabha National Institute, Kolkata, India
\item \Idef{org109}School of Physics and Astronomy, University of Birmingham, Birmingham, United Kingdom
\item \Idef{org110}Secci\'{o}n F\'{\i}sica, Departamento de Ciencias, Pontificia Universidad Cat\'{o}lica del Per\'{u}, Lima, Peru
\item \Idef{org111}Shanghai Institute of Applied Physics, Shanghai, China
\item \Idef{org112}St. Petersburg State University, St. Petersburg, Russia
\item \Idef{org113}Stefan Meyer Institut f\"{u}r Subatomare Physik (SMI), Vienna, Austria
\item \Idef{org114}SUBATECH, IMT Atlantique, Universit\'{e} de Nantes, CNRS-IN2P3, Nantes, France
\item \Idef{org115}Suranaree University of Technology, Nakhon Ratchasima, Thailand
\item \Idef{org116}Technical University of Ko\v{s}ice, Ko\v{s}ice, Slovakia
\item \Idef{org117}Technische Universit\"{a}t M\"{u}nchen, Excellence Cluster 'Universe', Munich, Germany
\item \Idef{org118}The Henryk Niewodniczanski Institute of Nuclear Physics, Polish Academy of Sciences, Cracow, Poland
\item \Idef{org119}The University of Texas at Austin, Austin, Texas, United States
\item \Idef{org120}Universidad Aut\'{o}noma de Sinaloa, Culiac\'{a}n, Mexico
\item \Idef{org121}Universidade de S\~{a}o Paulo (USP), S\~{a}o Paulo, Brazil
\item \Idef{org122}Universidade Estadual de Campinas (UNICAMP), Campinas, Brazil
\item \Idef{org123}Universidade Federal do ABC, Santo Andre, Brazil
\item \Idef{org124}University of Cape Town, Cape Town, South Africa
\item \Idef{org125}University of Houston, Houston, Texas, United States
\item \Idef{org126}University of Jyv\"{a}skyl\"{a}, Jyv\"{a}skyl\"{a}, Finland
\item \Idef{org127}University of Liverpool, Liverpool, United Kingdom
\item \Idef{org128}University of Science and Techonology of China, Hefei, China
\item \Idef{org129}University of South-Eastern Norway, Tonsberg, Norway
\item \Idef{org130}University of Tennessee, Knoxville, Tennessee, United States
\item \Idef{org131}University of the Witwatersrand, Johannesburg, South Africa
\item \Idef{org132}University of Tokyo, Tokyo, Japan
\item \Idef{org133}University of Tsukuba, Tsukuba, Japan
\item \Idef{org134}Universit\'{e} Clermont Auvergne, CNRS/IN2P3, LPC, Clermont-Ferrand, France
\item \Idef{org135}Universit\'{e} de Lyon, Universit\'{e} Lyon 1, CNRS/IN2P3, IPN-Lyon, Villeurbanne, Lyon, France
\item \Idef{org136}Universit\'{e} de Strasbourg, CNRS, IPHC UMR 7178, F-67000 Strasbourg, France, Strasbourg, France
\item \Idef{org137}Universit\'{e} Paris-Saclay Centre d'Etudes de Saclay (CEA), IRFU, D\'{e}partment de Physique Nucl\'{e}aire (DPhN), Saclay, France
\item \Idef{org138}Universit\`{a} degli Studi di Foggia, Foggia, Italy
\item \Idef{org139}Universit\`{a} degli Studi di Pavia, Pavia, Italy
\item \Idef{org140}Universit\`{a} di Brescia, Brescia, Italy
\item \Idef{org141}Variable Energy Cyclotron Centre, Homi Bhabha National Institute, Kolkata, India
\item \Idef{org142}Warsaw University of Technology, Warsaw, Poland
\item \Idef{org143}Wayne State University, Detroit, Michigan, United States
\item \Idef{org144}Westf\"{a}lische Wilhelms-Universit\"{a}t M\"{u}nster, Institut f\"{u}r Kernphysik, M\"{u}nster, Germany
\item \Idef{org145}Wigner Research Centre for Physics, Hungarian Academy of Sciences, Budapest, Hungary
\item \Idef{org146}Yale University, New Haven, Connecticut, United States
\item \Idef{org147}Yonsei University, Seoul, Republic of Korea
\end{Authlist}
\endgroup
\end{document}